\documentclass
	[
		10 pt
	]
	{article}
\usepackage{tikz, pgf, pgfplots}
\usetikzlibrary
	{
		3d,
		arrows,											
		arrows.meta,
		automata,										
		backgrounds,
		bending,
		calc,
		chains,
		datavisualization.formats.functions,			
		decorations.pathmorphing,
		decorations.pathreplacing,
		decorations.text,
		fadings,
		fit,
		graphs,
		graphs.standard,
		matrix,
		positioning,									
		quantikz2,
		quotes,
		shadings,
		shadows.blur,
		shapes,
		trees
	}
\pgfplotsset{compat = newest}
\usepgflibrary
	{
		shapes.misc
	}
\usepgfplotslibrary
	{
		dateplot
	}
\usepackage{tikzpeople}
\usepackage{fontawesome5}

\usepackage{amsthm}
\usepackage{amssymb}
\usepackage[bb = boondox]{mathalfa}
\usepackage{dsfont}
\usepackage{newtxmath}
\usepackage{mathtools}
\usepackage{multicol}
\usepackage{braket}
\usepackage{physics}
\usepackage{empheq}

\usepackage[ compat = newest ]{yquant}

\usepackage{xcolor}
\usepackage{varwidth}
\usepackage[ skins, breakable, listings, theorems ] {tcolorbox}
\usepackage{graphicx}
\usepackage{xurl}                                                                                  %
\usepackage{hyperref}
\hypersetup
	{
		colorlinks,
		linkcolor = {WordRed!75!black},
		citecolor = {WordBlueDarker25},
		urlcolor = {WordAquaDarker50}
	}
\usepackage{colortbl}
\usepackage{float}
\usepackage{nicematrix}
\usepackage{cellspace}
\usepackage{makecell}
\usepackage{multirow}
\usepackage{caption}
\usepackage[ linesnumbered, tworuled, vlined ] {algorithm2e}
\usepackage{appendix}
\usepackage{enumitem}
\usepackage{siunitx}
\usepackage{xintexpr}
\usepackage{spreadtab}
\usepackage{framed}
\usepackage{svg}
\usepackage{lipsum}
\newcounter{mathseed}
\pgfmathsetseed{\arabic{mathseed}}
\pgfdeclarelayer{background}
\pgfsetlayers{background, main}
\pgfdeclaredecoration{irregular fractal line}{init}
{
	\state{init}[width=\pgfdecoratedinputsegmentremainingdistance]
	{
		\pgfpathlineto{\pgfpoint{random*\pgfdecoratedinputsegmentremainingdistance}{(random*\pgfdecorationsegmentamplitude-0.02)*\pgfdecoratedinputsegmentremainingdistance}}
		\pgfpathlineto{\pgfpoint{\pgfdecoratedinputsegmentremainingdistance}{0pt}}
	}
}
\def\tornpaper#1{%
	\ifthenelse{\isodd{\value{mathseed}}}
	{%
		\tikz
		{
			\node[inner sep = 1em] (A) {#1};		
			\begin{pgfonlayer}{background}			
				\fill[paper]						
				\pgfextra{\pgfmathsetseed{\arabic{mathseed}}\addtocounter{mathseed}{1}}%
				{decorate[irregular cloudy border]{decorate{decorate{decorate{decorate[ragged border]{
										(A.north west) -- (A.north east)
				}}}}}}
				-- (A.south east)
				\pgfextra{\pgfmathsetseed{\arabic{mathseed}}}%
				{decorate[irregular spiky border]{decorate{decorate{decorate{decorate[ragged border]{
										-- (A.south west)
				}}}}}}
				-- (A.north west);
			\end{pgfonlayer}
		}
	}
	{%
		\tikz{
			\node[inner sep=1em] (A) {#1};  
			\begin{pgfonlayer}{background}  
				\fill[paper] 
				\pgfextra{\pgfmathsetseed{\arabic{mathseed}}\addtocounter{mathseed}{1}}%
				{decorate[irregular spiky border]{decorate{decorate{decorate{decorate[ragged border]{
										(A.north east) -- (A.north west)
				}}}}}}
				-- (A.south west)
				\pgfextra{\pgfmathsetseed{\arabic{mathseed}}}%
				{decorate[irregular cloudy border]{decorate{decorate{decorate{decorate[ragged border]{
										-- (A.south east)
				}}}}}}
				-- (A.north east);
		\end{pgfonlayer}}
	}
}
\definecolor{MyLightRed}{RGB}{244, 213, 245}
\definecolor{WordRed}{RGB}{255, 0, 102}
\definecolor{WordRedAccent5Lighter60}{HTML}{F5B5A7}
\definecolor{WordRedAccent5Darker25}{HTML}{B23214}
\definecolor{RedDarkLightest}{HTML}{ff0088}
\definecolor{RedDarkLight}{HTML}{ea005f}
\definecolor{RedPurple}{HTML}{aa007f}
\definecolor{Purple}{HTML}{911146}
\definecolor{PurpleDark}{RGB}{102, 0, 102}
\definecolor{WordLightGreen}{RGB}{140, 214, 192}
\definecolor{WordGreen}{RGB}{0, 176, 80}
\definecolor{GreenLightest}{HTML}{00ffa0}
\definecolor{GreenLighter1}{HTML}{00b383}
\definecolor{GreenLighter2}{HTML}{00aa7f}
\definecolor{GreenDark}{HTML}{225522}
\definecolor{GreenTeal}{HTML}{008080}
\definecolor{WordIceBlue}{RGB}{223, 227, 229}
\definecolor{MyVeryLightBlue}{RGB}{211, 245, 247}
\definecolor{WordBlueVeryLight}{RGB}{0, 176, 240}
\definecolor{WordBlueLight}{RGB}{0, 112, 192}
\definecolor{WordBlueDark}{RGB}{46, 116, 181}
\definecolor{WordBlueDarker}{RGB}{31, 78, 121}
\definecolor{WordBlueDarker25}{RGB}{54, 96, 146}
\definecolor{WordBlueDarker50}{RGB}{36, 64, 98}
\definecolor{WordBlueDarkest}{RGB}{0, 32, 96}
\definecolor{WordBlue}{RGB}{19, 65, 99}
\definecolor{MyBlue}{RGB}{0, 64, 128}
\definecolor{MyDarkBlue}{RGB}{0, 51, 102}
\definecolor{BlueVeryDark}{HTML}{222255}
\definecolor{MagentaVeryLight}{RGB}{178, 162, 201}
\definecolor{MagentaLighter}{RGB}{161, 106, 221}
\definecolor{MagentaLight}{RGB}{128, 100, 162}
\definecolor{MagentaDark}{RGB}{106, 65, 152}
\definecolor{MagentaVeryDark}{RGB}{97, 75, 128}
\definecolor{WordAquaLighter80}{RGB}{218, 238, 243}
\definecolor{WordAquaLighter60}{RGB}{183, 222, 232}
\definecolor{WordAquaLighter40}{RGB}{146, 205, 220}
\definecolor{WordAquaDarker25}{RGB}{49, 134, 155}
\definecolor{WordAquaAccent2Darker25}{HTML}{398E98}
\definecolor{WordAquaDarker50}{RGB}{33, 89, 103}
\definecolor{WordVeryLightTeal}{RGB}{223, 236, 235}
\definecolor{WordLightTeal}{RGB}{160, 199, 197}
\definecolor{WordDarkTealLighter80}{RGB}{207, 223, 234}
\definecolor{WordDarkTeal}{RGB}{72, 123, 119}
\definecolor{WordDarkerTeal}{RGB}{48, 82, 80}
\definecolor{WordTurquoiseLighter80}{RGB}{209, 238, 249}
\definecolor{WordGoldAccent1Lighter40}{HTML}{FFDF6A}
\definecolor{WordGoldAccent1Darker25}{HTML}{C49A00}
\definecolor{Brown}{HTML}{666633}
\definecolor{WordOrangeAccent2Lighter60}{HTML}{FCD3A4}
\definecolor{WordOrangeAccent4Lighter60}{HTML}{F7C5A1}

\newtheorem{definition}{Definition}[section]

\usepackage
	[
		a4paper,
		left = 2.5cm,
		right = 2.5 cm,
		top = 2.5cm,
		bottom = 3.0 cm
	]
	{geometry}
\title
	{
		A novel $2$ \& $3$ player scheme for Quantum Direct Communication
	}

\usepackage{orcidlink}
\newcommand{\orcidicon}[1]{\href{https://orcid.org/#1}{\includegraphics[height=\fontcharht\font`\B]{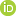}}}

\author
{
	Theodore Andronikos$^1$\orcidicon{0000-0002-3741-1271}
	and
	Alla Sirokofskich$^2$\\
	\\
	$^1$ \ Department of Informatics, Ionian University, \\
	7 Tsirigoti Square, 49100 Corfu, Greece; \\
	andronikos@ionio.gr \\
	$^2$ \ Department of History and Philosophy of Sciences, \\
	National and Kapodistrian University of Athens, \\
	Athens 15771, Greece; \\
	asirokof@math.uoa.gr
}
\begin{document}

\maketitle

\begin{abstract}
	This paper introduces two information-theoretically secure protocols that achieve quantum secure direct communication between Alice and Bob in the first case, and among Alice, Bod and Charlie in the second case. Both protocols use the same novel method to embed the secret information in the entangled compound system. The way of encoding the information is the main novelty of this paper and the distinguishing feature compared to previous works in this field. The most critical advantage of this method is that it is easily scalable and extensible because it can be seamlessly generalized to a setting involving three, or even more, players, as demonstrated with the second protocol. This trait can be extremely beneficial when many spatially separated players posses only part the secret information that must be combined and transmitted to Alice, so that she can obtain the complete secret. Using the three player protocol, this task can be achieved in one go, without the need to apply a typical QSDC protocol twice, where Alice first receives Bob's information and afterwards Charlie's information. The proposed protocol doesn't require pre-shared keys or quantum signatures, making it less complicated and more straightforward. Finally, by employing only standard CNOT and Hadamard gates, it offers the important practical advantage of being implementable on contemporary quantum computers, especially in view of the coming era of quantum distributed computing.
\end{abstract}
\section{Introduction} \label{sec: Introduction}

It is hardly necessary to advocate the importance of privacy and security for every aspect of our life as individuals. Indeed, privacy is a constitutional right that must be respected and protected under all circumstances. This, in turn, has advanced the design and implementation of technical tools that ensure the security of our digital data. Devising bulletproof algorithms and protocols that protect our privacy from unauthorized access is a major trend in current research. This, however, may not be as easy as it sounds. The reason is that we have just entered a new scientific era, the quantum era, which brings the promise of unprecedented computational power. This, unharnessed so far, power offers new algorithms that can, potentially compromise the security offered by established classical methods. Two iconic examples that help drive this point home, are Shor’s \cite{Shor1994} and Grover's \cite{Grover1996} algorithms. Shor’s algorithm can factorize large numbers in polynomial time and its practical implementation is bound to threaten public key cryptosystems. Grover’s algorithm speeds up unordered search and may also be used to attack symmetric key cryptosystems like AES.

Up to this day, there exist no quantum computers powerful enough to threaten the classical status quo. However, this will probably change sooner than initially anticipated, if one judges by the impressive progress that has been achieved lately. As evidence of this recent rapid advancement we mention IBM's 127-qubit Eagle \cite{IBMEagle2021}, the 433-qubit Osprey \cite{IBMOsprey2022}, the 1,121-qubit Condor \cite{IBMCondor2023}, and the newest and most powerful yet R2 Heron \cite{IBMHeron2024}. HENCE, It seems prudent, if not imperative, to find ways to seriously upgrade our algorithms and protocols, before they become a liability to our security infrastructure. The enormous effort to come up with a robust solution, has led to the creation of two new scientific fields, the field of post-quantum or quantum-resistant cryptography and the field of quantum cryptography. The former, is actually an incremental evolution of the current state of affairs \cite{chen2016report, alagic2019status, alagic2020status, alagic2022status}, reducing security issues to carefully chosen computationally hard problems, an approach that has been vindicated so far. The latter, quantum cryptography, relies on the laws of nature, such as entanglement, monogamy of entanglement, the no-cloning theorem, and nonlocality to ensure ironclad security. Quantum cryptography advocates exploiting the unique and powerful quantum phenomena to design new secure protocols for a plethora of critical applications, such as key distribution \cite{Bennett1984, Ekert1991, Gisin2004, inoue2002differential, guan2015experimental, waks2006security, Ampatzis2021}, secret sharing \cite{Hillery1999, Ampatzis2022, Ampatzis2023, Andronikos2024b}, quantum teleportation \cite{Bennett1993}, cloud storage \cite{attasena2017secret, ermakova2013secret} and blockchain \cite{cha2021blockchain, Sun2020, Qu2023}.

The most significant obstacle to the potential use of quantum computers for problems at the industrial scale is the requirement to scale them up. By now, it is obvious that scientific advancements and the removal of many technological barriers are required to advance the scale of quantum computers beyond the NISQ level. The development of distributed quantum computing systems is, in our opinion, currently the most promising way to overcome the scaling issue. A distributed quantum computer is made up of a network of quantum computing nodes that can transmit both classical and quantum data and each have a specific number of qubits available for processing. Considering that quantum and classical computing differ fundamentally, designing networked quantum computers poses special difficulties that are not present in classical networks. Towards this goal, there have been lately significant technological advancements in terms of hardware \cite{Photonic2024, NuQuantum2024} and design concepts \cite{Cacciapuoti2024, Illiano2024} that imply that distributed quantum computers will arrive sooner than anticipated. Thus, in a very concrete sense, we can say that the era of distributed quantum computing has arrived.

In the seminal paper \cite{Long2002}, the authors proposed a protocol for Quantum Secure Direct Communication (QSDC for short). The characteristic trait of QSDC, which distinguishes it from key distribution that establishes a common random key between two parties, is that QSDC transmits information directly through the quantum channel and without using an existing key. Furthermore, the classical channel is employed only for detection purposes and not for transmitting information necessary to decipher the secret message. The intended recipient must be able to uncover the secret information after receiving the quantum states via the quantum channel. Finally, any eavesdropper must be detected, without being allowed to compromise the secret. Almost immediately, in 2003 the researchers in \cite{Deng2003} introduced the influential two-step QSDC protocol. Later, \cite{Deng2004} presented a QSDC protocol using single photons, \cite{Wang2005} gave a protocol based on superdense coding, and \cite{Wang2005a} proposed the first QSDC protocol with multipartite entanglement. Since then, progress in this are has been non-stop. For a though and comprehensive review of the current state of the field, we refer the reader to the very recent \cite{Pan2023} and \cite{Pan2024}.

In this work, we initially introduce a new protocol, called $2$PSQDC, for quantum secure direct communication between two entities. Subsequently, the $2$PSQDC is generalized in a intuitive and straightforward manner, so as to provide for quantum secure direct communication among three entities. The resulting protocol, which is called $3$PSQDC, can be seamlessly generalized to an arbitrary number of entities. We present our protocols as games, involving the usual cast of Alice, Bob and Charlie. Hopefully, the pedagogical nature of games will make the presentation of the technical concepts easier to follow. Quantum games, from their inception in 1999 \cite{Meyer1999,Eisert1999}, have known great acceptance since quantum strategies are sometimes superior to classical ones \cite{Andronikos2018,Andronikos2021,Andronikos2022a}. The famous prisoners' dilemma game provides such the most prominent example \cite{Eisert1999}, which also applies to other abstract quantum games \cite{Giannakis2015a,Koh2024}. Many classical systems can be quantized, even political structures as was shown in \cite{Andronikos2022}. While on the subject of games on unconventional environments, let us mention that games in biological systems have attracted significant attention \cite{Theocharopoulou2019,Kastampolidou2020a,Kostadimas2021}. It is interesting to observe that biosystems may give rise to biostrategies superior compared to the classical ones, even in the Prisoners' Dilemma iconic game \cite{Kastampolidou2020,Kastampolidou2021,Papalitsas2021,Kastampolidou2023,Adam2023}.

\textbf{Contribution}. This paper presents two protocols that achieve quantum secure direct communication between Alice and Bob in the first case, and among Alice, Bod and Charlie in the second case. Both protocols, which are proven to be information-theoretically secure, use the same idea, i.e., embedding the secret information into the global state of the entangled composite system via a unitary transform that uses the inner product modulo $2$ operation. This way of encoding the information is the main novelty of this paper that distinguishes it from the previous works in the field. The advantage of this method is that it is seamlessly extensible and can be generalized to a setting involving three, or even more, players, as demonstrated with the $3$PSQDC protocol. This last case, is not only useful, but often necessary, when two spatially separated players posses only part the secret information that must be combined and transmitted to Alice in order for her to reveal the complete secret. Using the $3$PSQDC protocol, this task can be achieved in one go, without the need to apply a typical QSDC protocol twice, where Alice first receives Bob's information and afterwards Charlie's information. As we advocated above, we believe that we are entering the era of quantum distributed computing. The proposed protocols are designed to be implemented on contemporary quantum computers, since they are characterized by uniformity and simplicity, use exclusively standard CNOT and Hadamard gates, and rely on EPR pairs and $\ket{ GHZ_{ 3 } }$ triplets. Lastly, an additional advantage promoting their practicality is the fact that they don't require pre-shared keys or quantum signatures.

\subsection*{Organization} \label{subsec: Organization}

The paper is organized as follows. Section \ref{sec: Introduction} contains an introduction to the subject along with bibliographic pointers to related works. Section \ref{sec: Background & Notation} explains the underlying theory required for the understanding of the protocols. Section \ref{sec: The $2$PSQDC Protocol} provides a detailed presentation of the $2$PSQDC protocol that tackles information transmission from Alice to Bob. Section {sec: The $3$PSQDC Protocol} contains a formal presentation of the $3$PSQDC protocol that, in addition to Bob and Alice, also involves Charlie. Finally, Section {sec: Discussion and Conclusions} gives a brief summary of this work, and outlines directions for future research.

\section{Background \& notation} \label{sec: Background & Notation}

\subsection{$\ket{ \Phi^{ + } }$ EPR pairs} \label{subsec: Phi^{ + } EPR Pairs}

There are certain properties of quantum physics that are quite strange, in the sense that they have no analogue in classical physics and even contradict our everyday intuition. Undoubtedly, entanglement falls into this category. This strange phenomenon is also a source of great potential, as it seems to be one of the keys for achieving things that are difficult or impossible in the classical world. Technically, entanglement appears in composite quantum systems, consisting of at least two subsystems, which can be, and usually are, spatially separated. In mathematical terms, a composite system is entangled, if its state must be described as a linear combination of two or more product states of its subsystems. Bell states, also referred to as EPR pairs, provide the most well-known example of maximal entanglement for a two-qubit system. There are four Bell states expressed as shown below (see \cite{Nielsen2010}). We use the subscripts $A$ and $B$ to make explicitly clear that the first qubit belongs to Alice and the second to Bob.

\begin{tcolorbox}
	[
		grow to left by = 0.000 cm,
		grow to right by = 0.000 cm,
		colback = white,			
		enhanced jigsaw,			
		sharp corners,
		toprule = 0.100 pt,
		bottomrule = 0.100 pt,
		leftrule = 0.100 pt,
		rightrule = 0.100 pt,
		sharp corners,
		center title,
		fonttitle = \bfseries
	]
	\begin{minipage}[b]{0.450 \textwidth}
		\begin{align}
			\label{eq: Bell State Phi +}
			\ket{ \Phi^{ + } }
			=
			\frac { \ket{ 0 }_{ A } \ket{ 0 }_{ B } + \ket{ 1 }_{ A } \ket{ 1 }_{ B } } { \sqrt{ 2 } }
		\end{align}
	\end{minipage} 
	\hfill
	\begin{minipage}[b]{0.450 \textwidth}
		\begin{align}
			\label{eq: Bell State Phi -}
			\ket{ \Phi^{ - } }
			=
			\frac { \ket{ 0 }_{ A } \ket{ 0 }_{ B } - \ket{ 1 }_{ A } \ket{ 1 }_{ B } } { \sqrt{ 2 } }
		\end{align}
	\end{minipage}
	\begin{minipage}[b]{0.450 \textwidth}
		\begin{align}
			\label{eq: Bell State Psi +}
			\ket{ \Psi^{ + } }
			=
			\frac { \ket{ 0 }_{ A } \ket{ 1 }_{ B } + \ket{ 1 }_{ A } \ket{ 0 }_{ B } } { \sqrt{ 2 } }
		\end{align}
	\end{minipage} 
	\hfill
	\begin{minipage}[b]{0.450 \textwidth}
		\begin{align}
			\label{eq: Bell State Psi -}
			\ket{ \Psi^{ - } }
			=
			\frac { \ket{ 0 }_{ A } \ket{ 1 }_{ B } - \ket{ 1 }_{ A } \ket{ 0 }_{ B } } { \sqrt{ 2 } }
		\end{align}
	\end{minipage}
\end{tcolorbox}

One of the critical advantages of quantum entanglement is that when one qubit of the pair gets measured, the other immediately collapses to the corresponding state, irrespective of the distance between them. It is precisely this celebrated trait of quantum entanglement that is utilized in quantum cryptographic protocols, e.g., for key distribution, secret sharing, etc. Obviously, to implement a complex protocol, a sequence of EPR pairs is required. In the $2$PSQDC protocol, we shall be using $\ket{ \Phi^{ + } }$ pairs. The mathematical description of a sequence of $m$ $\ket{ \Phi^{ + } }$ pairs is

\begin{align}
	\label{eq: m-Fold Phi + Pairs}
	\ket{ \Phi^{ + } }^{ \otimes m }
	&=
	\frac { 1 } { \sqrt{ 2^m } }
	\sum_{ \mathbf{ x } \in \mathbb{ B }^{ m } }
	\ket{ \mathbf{ x } }_{ A }
	\ket{ \mathbf{ x } }_{ B }
	\ ,
\end{align}

where $\mathbb{ B } = \{ 0, 1 \}$.

\subsection{$\ket{ GHZ_{ 3 } }$ triplets} \label{subsec: GHZ_{ 3 } Triplets}

The phenomenon of entanglement appears not only in two qubit systems, but also in general multipartite system. For a composite system consisting of three or more qubits, one of the most well-known and studied types of maximal entanglement is the so-called GHZ state. In the $3$PSQDC protocol we shall employ triplets of qubits in the $\ket{ GHZ_{ 3 } }$ state. The latter is expressed mathematically by the equation \eqref{eq: GHZ 3 State}. As in the case of the $\ket{ \Phi^{ + } }$ pairs, subscripts $A$, $B$ and $C$ are used to make clear that the first qubit belongs to Alice, the second to Bob and the third to Charlie.

\begin{align}
	\label
	{eq: GHZ 3 State}
	\ket{ GHZ_{ 3 } }
	=
	\frac
	{
		\ket{ 0 }_{ A }
		\ket{ 0 }_{ B }
		\ket{ 0 }_{ C }
		+
		\ket{ 1 }_{ A }
		\ket{ 1 }_{ B }
		\ket{ 1 }_{ C }
	}
	{ \sqrt{ 2 } }
	\ .
\end{align}

A single $\ket{ GHZ_{ 3 } }$ triplet will not suffice for the execution of the $3$PSQDC protocol; $m$ such triplets will be required. A system comprised of $m$ $\ket{ GHZ_{ 3 } }$ triplets is described by the next formula (for its detailed derivation we refer to \cite{Ampatzis2022} and \cite{Ampatzis2023}):

\begin{align}
	\label{eq: m-Fold GHZ 3 States}
	\ket{ GHZ_{ 3 } }^{ \otimes m }
	=
	\frac { 1 } { \sqrt{ 2^{ m } } }
	\sum_{ \mathbf{ x } \in \mathbb{ B }^{ m } }
	\ket{ \mathbf{ x } }_{ A }
	\ket{ \mathbf{ x } }_{ B }
	\ket{ \mathbf{ x } }_{ C }
	\ .
\end{align}

In formulae \eqref{eq: m-Fold Phi + Pairs} and \eqref{eq: m-Fold GHZ 3 States}, the notation $\mathbf{ x } \in \mathbb{ B }^{ m }$ means that the bit vector $\mathbf{ x }$ ranges through all the $2^{ m }$ binary representations of the basis kets. In accordance to what we mentioned before, $\ket{ \mathbf{ x } }_{ A }$, $\ket{ \mathbf{ x } }_{ B }$ and $\ket{ \mathbf{ x } }_{ C }$ correspond to the basis states of Alice, Bob and Charlie's quantum registers, respectively.

Existing quantum computers based on the circuit model can trivially produce the four Bell states.
Similarly, it is easy in principle to construct quantum circuits that produce general $\ket{ GHZ_{ n } }$ states. As a matter of fact there is a methodology for constructing efficient general GHZ circuits \cite{Cruz2019}, in the sense that it takes $\lg n$ steps to produce the given state. Although, for large $n$ there are practical difficulties in preparing and maintaining $\ket{ GHZ_{ n } }$ states, for the $\ket{ GHZ_{ 3 } }$ states used in the implementation of the $3$PSQDC protocol, things are quite manageable.

Apart from $\ket{ \Phi^{ + } }$ pairs and $\ket{ GHZ_{3} }$ triplets, we use the well-known states $\ket{+}$ and $\ket{-}$. For completeness, we provide their definitions.

\begin{tcolorbox}
	[
		grow to left by = 0.000 cm,
		grow to right by = 0.000 cm,
		colback = white,			
		enhanced jigsaw,			
		sharp corners,
		toprule = 0.100 pt,
		bottomrule = 0.100 pt,
		leftrule = 0.100 pt,
		rightrule = 0.100 pt,
		sharp corners,
		center title,
		fonttitle = \bfseries
	]
	\begin{minipage}[b]{0.450 \textwidth}
		\begin{align}
			\label{eq: Ket +}
			\ket{ + }
			=
			H
			\ket{ 0 }
			=
			\frac
			{ \ket{ 0 } + \ket{ 1 } }
			{ \sqrt{ 2 } }
		\end{align}
	\end{minipage} 
	\hfill
	\begin{minipage}[b]{0.450 \textwidth}
		\begin{align}
			\label{eq: Ket -}
			\ket{ - }
			=
			H
			\ket{ 1 }
			=
			\frac
			{ \ket{ 0 } - \ket{ 1 } }
			{ \sqrt{ 2 } }
		\end{align}
	\end{minipage}
\end{tcolorbox}

Another useful formula proved in textbooks such as \cite{Nielsen2010, Mermin2007, Wong2022}, which will be applied in the explanation of the protocol, is the following

\begin{align}
	\label{eq: Hadamard m-Fold Ket x}
	H^{ \otimes m }
	\ket{ \mathbf{ x } }
	&=
	\frac
	{ 1 }
	{ \sqrt{ 2^{ m } } }
	\sum_{ \mathbf{ z } \in \mathbb{ B }^{ m } }
	( - 1 )^{ \mathbf{ z \cdot x } }
	\ket{ \mathbf{ z } }
	\ .
\end{align}

Finally, we mention that, as a rule, quantum measurements are performed with respect to the computational basis $\{ \ket{ 0 }, \ket{ 1 } \}$. However, occasionally, measurements are also made in the Hadamard basis $\{ \ket{ + }, \ket{ - } \}$; in such cases it is explicitly clarified.

\subsection{Inner product modulo $2$} \label{subsec: Inner Product Modulo $2$}

In this work, we follow the typical convention of writing bit vectors $\mathbf{ x } \in \mathbb{ B }^{ m }$ in boldface. A bit vector $\mathbf{ x }$ of length $m$ is a sequence of $m$ bits $\mathbf{ x } = x_{ m - 1 } \dots x_{ 0 }$. The zero bit vector is designated by $\mathbf{ 0 } = 0 \dots 0$. Given two bit vectors $\mathbf{ x }, \mathbf{ y } \in \mathbb{ B }^{ m }$, where $\mathbf{ x } = x_{ m - 1 } \dots x_{ 0 }$ and $\mathbf{ y } = y_{ m - 1 } \dots y_{ 0 }$, we define the \emph{inner product modulo} $2$, denoted by $\mathbf{ x \cdot y }$, as

\begin{align}
	\label{eq: Inner Product Modulo $2$}
	\mathbf{ x \cdot y }
	\coloneq
	x_{ n - 1 } y_{ n - 1 }
	\oplus
	\dots
	\oplus
	x_{ 0 } y_{ 0 }
	\ .
\end{align}

In the above formula, $\coloneq$ stands for ``is defined as,'' and $\oplus$ stands for addition modulo $2$. The operation inner product modulo $2$ exhibits a very useful property. If $\mathbf{ c } \in \mathbb{ B }^{ m }$ is different from $\mathbf{ 0 }$, then for half of the elements $\mathbf{ x } \in \mathbb{ B }^{ m }$, $\mathbf{c} \cdot \mathbf{ x }$ is $0$ and for the other half, $\mathbf{ c } \cdot \mathbf{ x }$ is $1$. Obviously, if $\mathbf{ c } = \mathbf{ 0 }$, then for all $\mathbf{ x } \in  \mathbb{ B }^{ m }$, $\mathbf{ c } \cdot \mathbf{ x } = 0$. For easy reference, we call this property the Characteristic Inner Product (CIP) property \cite{Andronikos2023b}.

\begin{align}
	\label{eq: The Characteristic Inner Product Modulo $2$ Property}
	\text{ If } \mathbf{ c } \neq \mathbf{ 0 },
	\text{ there are }
	&2^{ m - 1 } \text{ bit vectors } \mathbf{ x } \in \mathbb{ B }^{ m },
	\text{ such that } \mathbf{ c } \cdot \mathbf{ x } = 0, \text{ and }
	\nonumber \\
	&2^{ m - 1 } \text{ bit vectors } \mathbf{ x } \in \mathbb{ B }^{ m },
	\text{ such that } \mathbf{ c } \cdot \mathbf{ x } = 1 \ .
\end{align}

It will also be expedient to extend the operation of addition modulo $2$ to bitwise addition modulo $2$ between bit vectors. Given two bit vectors $\mathbf{ x }, \mathbf{ y } \in \mathbb{ B }^{ m }$, where $\mathbf{ x } = x_{ m - 1 } \dots x_{ 0 }$ and $\mathbf{ y } = y_{ m - 1 } \dots y_{ 0 }$, we define their \emph{bitwise addition modulo} $2$, denoted by $\mathbf{ x \oplus y }$, as

\begin{align}
	\label{eq: Bitwise Addition Modulo $2$}
	\mathbf{ x }
	\oplus
	\mathbf{ y }
	\coloneq
	( x_{ n - 1 } \oplus y_{ n - 1 } )
	\dots
	( x_{ 0 } \oplus y_{ 0 } )
	\ .
\end{align}

We use the same symbol $\oplus$ for the operation of addition modulo $2$ between bits, and for the operation of bitwise addition modulo $2$ between bit vectors, since the context will help prevent any confusion.

\section{The $2$PSQDC protocol} \label{sec: The $2$PSQDC Protocol}

In this section we present the first Quantum Secure Direct Communication (QSDC for now on) for the simpler case of two players. The $2$ player scheme for quantum secure direct communication, refereed to as $2$PSQDC, is designed to allow one party to communicate with a spatially separated second party securely and directly using only the quantum channel. To emphasize that it is designed to enable the secure direct communication of two players, the protocol is called $2$PSQDC. To enhance its gamelike presentation, we call the two parties Alice and Bob. Alice is the player having the initiative and intending to send some information to Bob. The setting is completed by the notorious Eve, a cunning adversary that attempts to steal any information possible. The major advantage that quantum protocols exhibit over classical ones is that communication through the quantum channel involves an array of unique features, such as the no-cloning theorem \cite{wootters1982single}, the monogamy of entanglement \cite{coffman2000distributed}, and nonlocality \cite{brunner2014bell}, that can be used to inhibit Eve. Before we proceed to the presentation of the $2$PSQDC protocol in earnest, we first clarify some important subtle points in the next subsection.

\subsection{The role of the classical channel \& defensive measures} \label{subsec: The Role of the Classical Channel & Defensive Measures}

The fundamental advantage of the QSDC scheme is the transmission of information exclusively through the quantum channel. However, it is important to emphasize that a classical authenticated channel is still necessary for QSDC to enforce security measures, such as eavesdropping detection, and other important tasks such as error correction.

The defensive capabilities of quantum protocols are improved by a few standard methods. The \emph{decoy technique} is one such method that is essential to our setup and is intended for \emph{eavesdropping detection}, i.e., revealing the existence of a potential eavesdropper. This method, in addition to the entangled information carrying tuples, creates extra decoy tuples, which are produced and distributed by the trusted party. These decoys are inserted at random locations in the transmission sequence(s). While Eve tampers with the transmission sequences unaware of the locations of the decoys, she will cause errors that the real players can identify. Since existing literature has extensively studied this technique  \cite{Deng2008,Yang2009,Tseng2011,Chang2013, Hung2016,Ye2018,Wu2021,Pan2023,Hou2024,Pan2024}, and this work has nothing further to add, we shall henceforth take for granted the implementation of such method, without providing superfluous details.

Needless to mention that, particularly in the case of entanglement-based protocols, entanglement is a prerequisite for their successful implementation. Verifying the existence of entanglement means that the protocol can continue to accomplish its stated goal. In contrast, without entanglement, the protocol is bound to fail. Therefore, it is essential to develop a reliable test for \emph{entanglement validation}. Absence of entanglement may be caused by an adversary's tampering or a noisy quantum channel, among other things. In such a situation, the only reasonable course of action is to terminate the protocol and start over after the necessary corrective actions have been taken. The significance of entanglement validation has led to a great deal of research in the literature. Our protocol follows the advanced methodologies described in previous studies, including \cite{Neigovzen2008,Feng2019,Wang2022a,Yang2022,Qu2023,Ikeda2023c}.

Moreover, this work is adheres to the computer scientist perspective, and, thus, both our protocols are described under the assumption of ideal quantum channels that don't take into account noise and loss.

\subsection{Entanglement distribution phase} \label{subsec: $2$PSQDC Entanglement Distribution Phase}

The $2$PSQDC protocol evolves in phases. Initially, during the entanglement distribution phase, Alice, or a third trusted source, prepares $\ket{ \Phi^{ + } }$ pairs. As customary, we assume the existence of a trusted quantum source, which may not necessarily be Alice, that is responsible for this task. In any event, the produced $\ket{ \Phi^{ + } }$ pairs are evenly shared between Alice and Bob. The precise pattern of the entanglement distribution follows the more general $r$-Uniform Distribution Scheme outlined in Definition \ref{def: Uniform Entanglement Distribution Scheme} (see \cite{Andronikos2024b}), where $r = 2$ for the $2$PSQDC protocol and $r = 3$ for the subsequent $3$PSQDC protocol.

\begin{definition} [Uniform Distribution Scheme] \label{def: Uniform Entanglement Distribution Scheme}
	The $r$-Uniform Distribution Scheme asserts that:
\end{definition}
\begin{itemize}
	\item	
	there are $r$ players and each player is endowed with a $p$-qubit register, and
	\item	
	the qubits in the $j^{ th }$ position, $0 \leq j \leq p - 1$, of these quantum registers are entangled in the $\ket{ GHZ_{ r } }$ state, or in the $\ket{ \Phi^{ + } }$ state when $r = 2$.
\end{itemize}

The situation after the preparation of the $\ket{ \Phi^{ + } }$ pairs and just before the entanglement distribution is visualized in Figure \ref{fig: Alice Creates $m$ Phi^{ + } Pairs}. At the end of the distribution phase, both Alice and Bob have in their own quantum registers $m$ qubits each. Their registers are correlated because Alice and Bob's corresponding qubits are entangled in the state $\ket{ \Phi^{ + } }$. The setup at this point is described in Figure \ref{fig: Alice Shares $m$ Phi^{ + } Pairs with Bob}. Note that in the above setup we have deliberately omitted the details about the extra EPR pairs that are required for eavesdropping detection and entanglement validation. This is done so as to facilitate the understanding of the $2$PSQDC protocol without clouting the presentation with extra technicalities.

To complete this phase, Alice and Bob conduct the eavesdropping detection and entanglement validation tests with the help of the classical channel. If these tests are successful, they proceed to the secret embedding phase. If not, then their communication has been compromised by Eve, and, so, they abort the protocol.


\begin{tcolorbox}
	[
		grow to left by = 0.000 cm,
		grow to right by = 0.000 cm,
		colback = WordTurquoiseLighter80!12,			
		enhanced jigsaw,								
		sharp corners,
		toprule = 1.000 pt,
		bottomrule = 1.000 pt,
		leftrule = 0.100 pt,
		rightrule = 0.100 pt,
		sharp corners,
		center title,
		fonttitle = \bfseries
	]
	\begin{figure}[H]
		\begin{minipage} [ b ] { 0.475 \textwidth }
			\centering
			\begin{tikzpicture} [ scale = 0.500 ]
				\node
					[
						shade, top color = MagentaLighter, bottom color = black, rectangle, text width = 4.000 cm, align = center
					]
					(Label)
					{ \color{white} \textbf{$\ket{ \Phi^{ + } }$ Pairs Preparation} };
				\node
					[
						alice,
						shirt = RedPurple,
						scale = 1.500,
						anchor = center,
						below left = 1.000 cm and 0.000 cm of Label,
						label = { [ label distance = 0.000 cm ] north: \textbf{Alice} }
					]
					(Alice) { };
				\node
					[
						bob,
						scale = 1.500,
						shirt = GreenLighter2,
						anchor = center,
						below right = 1.000 cm and 0.000 cm of Label,
						label = { [ label distance = 0.000 cm ] north: \textbf{Bob} }
					]
					(Bob) { };
				\node
					[
						circle, shade, outer color = RedPurple!50, inner color = white, minimum size = 5.000 mm, below = 0.500 cm of Alice,
						label = { [ label distance = 0.000 cm ] west: $\ket{ \Phi^{ + } }_{ m - 1 }$ }
					]
					(Phi+A-m-1) { };
				\node
					[
						circle, shade, outer color = GreenLighter2!50, inner color = white, minimum size = 5.000 mm, right = 7.000 mm of Phi+A-m-1,
					]
					(Phi+B-m-1) { };
				\draw
					[ MagentaDark, -, >=stealth, line width = 0.750 mm, decoration = coil, decorate ]
					(Phi+A-m-1) -- (Phi+B-m-1);
				\node
					[
						circle, shade, outer color = RedPurple!50, inner color = white, minimum size = 5.000 mm, below = 1.500 cm of Alice,
						label = { [ label distance = 0.000 cm ] west: $\ket{ \Phi^{ + } }_{ m - 2 }$ }
					]
					(Phi+A-m-2) { };
				\node
					[
					circle, shade, outer color = GreenLighter2!50, inner color = white, minimum size = 5.000 mm, right = 7.000 mm of Phi+A-m-2,
					]
					(Phi+B-m-2) { };
				\draw
					[ MagentaDark, -, >=stealth, line width = 0.750 mm, decoration = coil, decorate ]
					(Phi+A-m-2) -- (Phi+B-m-2);
				\node
					[
						minimum size = 5.000 mm, below = 2.000 cm of Alice,
					]
					(VDotsA)
					{ \vdots };
				\node
					[
						minimum size = 5.000 mm, right = 7.000 mm of VDotsA,
					]
					(VDotsB)
					{ \vdots };
				\node
					[
						circle, shade, outer color = RedPurple!50, inner color = white, minimum size = 5.000 mm, below = 3.000 cm of Alice,
						label = { [ label distance = 0.000 cm ] west: $\ket{ \Phi^{ + } }_{ 1 \phantom{ m - } }$ }
					]
					(Phi+A-1) { };
				\node
					[
						circle, shade, outer color = GreenLighter2!50, inner color = white, minimum size = 5.000 mm, right = 7.000 mm of Phi+A-1,
					]
					(Phi+B-1) { };
				\draw
					[ MagentaDark, -, >=stealth, line width = 0.750 mm, decoration = coil, decorate ]
					(Phi+A-1) -- (Phi+B-1);
				\node
					[
						circle, shade, outer color = RedPurple!50, inner color = white, minimum size = 5.000 mm, below = 4.000 cm of Alice,
						label = { [ label distance = 0.000 cm ] west: $\ket{ \Phi^{ + } }_{ 0 \phantom{ m - } }$ }
					]
					(Phi+A-0) { };
				\node
					[
						circle, shade, outer color = GreenLighter2!50, inner color = white, minimum size = 5.000 mm, right = 7.000 mm of Phi+A-0,
					]
					(Phi+B-0) { };
				\draw
					[ MagentaDark, -, >=stealth, line width = 0.750 mm, decoration = coil, decorate ]
					(Phi+A-0) -- (Phi+B-0);
				\node [ anchor = west, below = 0.100 cm of Phi+A-0 ] (PhantomNode1) { };
				\draw
					[ WordBlueVeryLight, <->, >=stealth, line width = 2.500 mm ]
					( $ (Alice.east) + ( 3.000 mm, 0.000 mm ) $ ) -- ( $ (Bob.west) - ( 3.000 mm, 0.000 mm ) $ ) node [ above = 0.250 cm, midway ] { \color{WordBlueDarkest} \textbf{Spatially Separated} };
			\end{tikzpicture}
			\caption{Alice creates the $\ket{ \Phi^{ + } }$ pairs she intends to share with Bob, who is at another spatial location.}
			\label{fig: Alice Creates $m$ Phi^{ + } Pairs}
		\end{minipage}
		\hfill
		\begin{minipage} [ b ] { 0.475 \textwidth }
			\centering
			\begin{tikzpicture} [ scale = 0.500 ]
				\node
					[
						shade, top color = MagentaLighter, bottom color = black, rectangle, text width = 4.000 cm, align = center
					]
					(Label)
					{ \color{white} \textbf{$\ket{ \Phi^{ + } }$ Pairs Distribution} };
				\node
					[
						alice,
						shirt = RedPurple,
						scale = 1.500,
						anchor = center,
						below left = 1.000 cm and 0.000 cm of Label,
						label = { [ label distance = 0.000 cm ] north: \textbf{Alice} }
					]
					(Alice) { };
				\node
					[
						bob,
						scale = 1.500,
						shirt = GreenLighter2,
						anchor = center,
						below right = 1.000 cm and 0.000 cm of Label,
						label = { [ label distance = 0.000 cm ] north: \textbf{Bob} }
					]
					(Bob) { };
				\node
					[
						circle, shade, outer color = RedPurple!50, inner color = white, minimum size = 5.000 mm, below = 0.500 cm of Alice,
						label = { [ label distance = 0.000 cm ] west: $\ket{ \Phi^{ + } }_{ m - 1 }$ }
					]
					(Phi+A-m-1) { };
				\node
					[
						circle, shade, outer color = GreenLighter2!50, inner color = white, minimum size = 5.000 mm, below = 0.500 cm of Bob,
					]
					(Phi+B-m-1) { };
				\draw
					[ MagentaDark, -, >=stealth, line width = 0.750 mm, decoration = coil, decorate ]
					(Phi+A-m-1) -- (Phi+B-m-1);
				\node
					[
						circle, shade, outer color = RedPurple!50, inner color = white, minimum size = 5.000 mm, below = 1.500 cm of Alice,
						label = { [ label distance = 0.000 cm ] west: $\ket{ \Phi^{ + } }_{ m - 2 }$ }
					]
					(Phi+A-m-2) { };
				\node
					[
						circle, shade, outer color = GreenLighter2!50, inner color = white, minimum size = 5.000 mm, below = 1.500 cm of Bob,
					]
					(Phi+B-m-2) { };
				\draw
					[ MagentaDark, -, >=stealth, line width = 0.750 mm, decoration = coil, decorate ]
					(Phi+A-m-2) -- (Phi+B-m-2);
				\node
					[
						minimum size = 5.000 mm, below = 2.000 cm of Alice,
					]
					(VDotsA)
					{ \vdots };
				\node
					[
						minimum size = 5.000 mm, below = 2.000 cm of Bob,
					]
					(VDotsB)
					{ \vdots };
				\node
					[
						circle, shade, outer color = RedPurple!50, inner color = white, minimum size = 5.000 mm, below = 3.000 cm of Alice,
						label = { [ label distance = 0.000 cm ] west: $\ket{ \Phi^{ + } }_{ 1 \phantom{ m - } }$ }
					]
					(Phi+A-1) { };
				\node
					[
						circle, shade, outer color = GreenLighter2!50, inner color = white, minimum size = 5.000 mm, below = 3.000 cm of Bob,
					]
					(Phi+B-1) { };
				\draw
					[ MagentaDark, -, >=stealth, line width = 0.750 mm, decoration = coil, decorate ]
					(Phi+A-1) -- (Phi+B-1);
				\node
					[
						circle, shade, outer color = RedPurple!50, inner color = white, minimum size = 5.000 mm, below = 4.000 cm of Alice,
						label = { [ label distance = 0.000 cm ] west: $\ket{ \Phi^{ + } }_{ 0 \phantom{ m - } }$ }
					]
					(Phi+A-0) { };
				\node
					[
						circle, shade, outer color = GreenLighter2!50, inner color = white, minimum size = 5.000 mm, below = 4.000 cm of Bob,
					]
					(Phi+B-0) { };
				\draw
					[ MagentaDark, -, >=stealth, line width = 0.750 mm, decoration = coil, decorate ]
					(Phi+A-0) -- (Phi+B-0);
				\node [ anchor = west, below = 0.100 cm of Phi+A-0 ] (PhantomNode1) { };
				\draw
					[ WordBlueVeryLight, <->, >=stealth, line width = 2.500 mm ]
					( $ (Alice.east) + ( 3.000 mm, 0.000 mm ) $ ) -- ( $ (Bob.west) - ( 3.000 mm, 0.000 mm ) $ ) node [ above = 0.250 cm, midway, text = black ] { \color{WordBlueDarkest} \textbf{Spatially Separated} };
			\end{tikzpicture}
			\caption
			{
				Alice shares the $\ket{ \Phi^{ + } }$ pairs with Bob by keeping the first and sending the second qubit.
			}
			\label{fig: Alice Shares $m$ Phi^{ + } Pairs with Bob}
		\end{minipage}
	\end{figure}
\end{tcolorbox}

\subsection{Secret embedding phase} \label{subsec: Secret Embedding Phase}

During this phase, Alice encodes the secret she intends to transmit to Bob. Let us assume that the secret information that Alice aims to convey to Bob is represented by the following bit vector $\mathbf{ s }$:

\begin{align}
	\label{eq: Secret Bit Vector}
	\mathbf{ s }
	=
	s_{ m - 1 }
	\dots
	s_{ 0 }
	\ .
\end{align}

To achieve this, Alice acts locally upon her register using the local quantum circuit outlined in Figure \ref{fig: $2$PSQDC Protocol Part 1}. Although Alice and Bob are spatially separated and they both operate via their local quantum circuits, the entanglement correlating their registers results in one composite system consisting of Alice and Bob's subsystems. In Figure \ref{fig: $2$PSQDC Protocol Part 1}, $AR$ and $BR$ designate Alice and Bob's entangled registers, respectively, and $AQ$ stands for Alice's qubit, initialized to state $\ket{ - }$. The subscripts $A$ and $B$ are used to distinguish between Alice and Bob's qubits and registers.

Alice embeds the secret into the state of the composite quantum system by using the unitary transform $U_{ A }$ on her quantum register. $U_{ A }$ is based on the function

\begin{align}
	\label{eq: Alice's Function}
	f_{ A } ( \mathbf{ x } )
	\coloneq
	\mathbf{ s }
	\cdot
	\mathbf{ x }
	\ .
\end{align}

The complete definition of $U_{ A }$ follows the typical rule given below

\begin{align}
	\label{eq: Alice's Transform}
	U_{ A }
	\colon
	\ket{ y }_{ A }
	\
	\ket{ \mathbf{ x } }_{ A }
	\rightarrow
	\ket{ y \oplus f_{ A } ( \mathbf{ x } ) }_{ A }
	\
	\ket{ \mathbf{ x } }_{ A }
	\ ,
\end{align}

where $\ket{ y }_{ A }$ and $\ket{ \mathbf{ x } }_{ A }$ represent the state of Alice's qubit $AQ$ and register $AR$, respectively. Taking into account equation \eqref{eq: Alice's Function} and the fact that $\ket{ y }_{ A } = \ket{ - }_{ A }$, equation \eqref{eq: Alice's Transform} becomes

\begin{align}
	\label{eq: Explicit Alice's Transform}
	U_{ A }
	\colon
	\ket{ - }_{ A }
	\
	\ket{ \mathbf{ x } }_{ A }
	\rightarrow
	( - 1 )^{ \mathbf{ s } \cdot \mathbf{ x } }
	\
	\ket{ - }_{ A }
	\
	\ket{ \mathbf{ x } }_{ A }
	\ .
\end{align}

The quantum circuit of Figure \ref{fig: $2$PSQDC Protocol Part 1} begins its operation in the initial state $\ket{ \psi_0 }$. By invoking \eqref{eq: m-Fold Phi + Pairs}, $\ket{ \psi_{ 0 } }$ can be expressed as

\begin{align}
	\label{eq: $2$PSQDC Part 1 Initial State}
	\ket{ \psi_{ 0 } }
	=
	\frac { 1 } { \sqrt{ 2^{ m } } }
	\sum_{ \mathbf{ x } \in \mathbb{ B }^{ m } }
	\
	\ket{ - }_{ A }
	\
	\ket{ \mathbf{ x } }_{ A }
	\
	\ket{ \mathbf{ x } }_{ B }
	\ .
\end{align}

\begin{tcolorbox}
	[
		enhanced,
		breakable,
		grow to left by = 0.000 cm,
		grow to right by = 0.000 cm,
		colback = MagentaLighter!07,			
		enhanced jigsaw,						
		sharp corners,
		toprule = 1.000 pt,
		bottomrule = 1.000 pt,
		leftrule = 0.100 pt,
		rightrule = 0.100 pt,
		sharp corners,
		center title,
		fonttitle = \bfseries
	]
	\begin{figure}[H]
		\centering
		\begin{tikzpicture} [ scale = 1.000 ]
			\begin{yquant}
				nobit AUX_B_0;
				nobit AUX_B_1;
				[ name = IBREG ] qubits { $BR$ \hspace{ 0.000 cm } } BR;
				nobit AUX_B_2;
				nobit AUX_B_3;
				nobit AUX_B_4;
				nobit AUX_B_5;
				nobit AUX_A_0;
				nobit AUX_A_1;
				nobit AUX_A_2;
				nobit AUX_A_3;
				[ name = IAREG ] qubits { $AR$ \hspace{ 0.000 cm } } AR;
				nobit AUX_A_4;
				qubit { $AQ$: \hspace{ 0.000 cm } $\ket{ - }$ } AQ;
				nobit AUX_A_5;
				nobit AUX_A_6;
				[ name = InitialState, WordBlueVeryLight, line width = 0.500 mm, label = { [ label distance = 0.200 cm ] north: Initial State } ]
				barrier ( - ) ;
				hspace { 0.500 cm } AR;
				[ draw = RedPurple, fill = RedPurple, radius = 0.700 cm ] box {\color{white} \Large \sf{U}$_{ A }$} (AR - AQ);
				[ name = SecretEmbedding, WordBlueVeryLight, line width = 0.500 mm, label = { [ label distance = 0.100 cm ] north: Secret Embedding } ]
				barrier ( - ) ;
				output { } BR;
				output { } AR;
				\node
				[
				bob,
				shirt = GreenLighter2,
				scale = 1.500,
				anchor = center,
				left = 0.250 cm of IBREG,
				label = { [ label distance = 0.000 cm ] north: \textbf{Bob} }
				]
				(Bob) { };
				\node
				[
				alice,
				shirt = RedPurple,
				scale = 1.500,
				anchor = center,
				left = 0.250 cm of IAREG,
				label = { [ label distance = 0.000 cm ] north: \textbf{Alice} }
				]
				(Alice) { };
				\node [ anchor = center, below = 1.750 cm of Alice ] (PhantomNode) { };
				\begin{scope} [ on background layer ]
					\node
					[ above right = 1.000 cm and 1.400 cm of Alice, rectangle, fill = WordAquaLighter60, text width = 6.500 cm, align = center, minimum height = 10.000 mm ] { \bf Spatially Separated };
					\draw
					[ MagentaDark, -, >=stealth, line width = 0.750 mm, decoration = coil, decorate ]
					( $ (IAREG.east) + ( 0.500 mm, 0.000 mm ) $ ) node [ circle, fill, minimum size = 1.500 mm ] () {} -- ( $ (IBREG.east) + ( 0.500 mm, 0.000 mm ) $ ) node [ circle, fill, minimum size = 1.500 mm ] () {};
				\end{scope}
				\node [ below = 3.100 cm ] at (InitialState) { $\ket{ \psi_{ 0 } }$ };
				\node [ below = 3.100 cm ] at (SecretEmbedding) { $\ket{ \psi_{ 1 } }$ };
			\end{yquant}
		\end{tikzpicture}
		\caption{The above figure is an abstract visualization of the quantum circuits employed by Alice and Bob in the secret embedding phase. Although the local quantum circuits are spatially separated, they constitute a compound system because they are linked by entanglement, indicated by the wavy magenta line connecting Alice and bob's registers.}
		\label{fig: $2$PSQDC Protocol Part 1}
	\end{figure}
\end{tcolorbox}

Alice acts on her quantum register with the unitary transform \eqref{eq: Explicit Alice's Transform} driving the composite system into the next state $\ket{ \psi_{ 1 } }$:

\begin{align}
	\label{eq: $2$PSQDC Part 1 Secret Embedding State}
	\ket{ \psi_{ 1 } }
	\overset { \eqref{eq: Explicit Alice's Transform} } { = }
	\frac { 1 } { \sqrt{ 2^{ m } } }
	\sum_{ \mathbf{ x } \in \mathbb{ B }^{ m } }
	\
	( - 1 )^{ \mathbf{ s } \cdot \mathbf{ x } }
	\
	\ket{ - }_{ A }
	\
	\ket{ \mathbf{ x } }_{ A }
	\
	\ket{ \mathbf{ x } }_{ B }
	\ .
\end{align}

At the end of this phase, Alice sends to Bob the $m$ qubits in her $AR$ register. This communication takes place exclusively through the quantum channel. Bob receives and places these qubits in a second register denoted by $BR_{ A }$.

Before Bob executes the final phase of the protocol, it is crucial that he and Alice perform once again the eavesdropping detection and entanglement validation tests. If both tests are successful, then Bob can confidently proceed to decipher the secret information. If not, then they surmise that the security has been breached and the protocol must be aborted.

\subsection{Secret decryption phase} \label{subsec: Secret Decryption Phase}

During this phase, Bob will complete the protocol and decipher Alice's secret locally, using the quantum circuit outlined in Figure \ref{fig: $2$PSQDC Protocol Part 2}. In addition to the quantum register $BR$, utilized in the circuit of Figure \ref{fig: $2$PSQDC Protocol Part 1}, Bob now also uses register $BR_{ A }$. These two registers are entangled because their corresponding qubits are $\ket{ \Phi^{ + } }$ pairs. By acting on both of them, Bob will uncover the secret bit vector $\mathbf{ s }$.

The circuit depicted in Figure \ref{fig: $2$PSQDC Protocol Part 2} starts its operation in the $\ket{ \psi_{ 1 } }$ state. Bob applies the $m$-fold Hadamard transform to both his registers $BR$ and $BR_{ A }$, driving the system to the next state $\ket{ \psi_{ 2 } }$. Note that in the rest of the computations we ignore Alice's local qubit $AQ$ that remains in state $\ket{ - }$, since it has served its intended purpose, which is to introduce the relative phase $( - 1 )^{ \mathbf{ s } \cdot \mathbf{ x } }$. Henceforth, to avoid any confusion, we shall use subscripts $BR_{ A }$ and $BR$ to refer to registers $BR_{ A }$ and $BR$, respectively.

\begin{tcolorbox}
	[
		enhanced,
		breakable,
		grow to left by = 0.000 cm,
		grow to right by = 0.000 cm,
		colback = MagentaLighter!07,			
		enhanced jigsaw,						
		sharp corners,
		toprule = 1.000 pt,
		bottomrule = 1.000 pt,
		leftrule = 0.100 pt,
		rightrule = 0.100 pt,
		sharp corners,
		center title,
		fonttitle = \bfseries
	]
	\begin{figure}[H]
		\centering
		\begin{tikzpicture} [ scale = 1.000 ]
			\begin{yquant}
				nobit AUX_B_0;
				[ name = IBREG ] qubits  { $BR$ \hspace{ 0.000 cm } } BR;
				nobit AUX_B_2;
				[ name = BOB ] nobit AUX_B_04;
				nobit AUX_B_6;
				[ name = IBREGA ] qubits { $BR_{ A }$ \hspace{ 0.000 cm } } BRA;
				nobit AUX_B_7;
				nobit AUX_B_8;
				[ name = Input, WordBlueVeryLight, line width = 0.500 mm, label = { [ label distance = 0.200 cm ] north: Input } ]
				barrier ( - ) ;
				hspace { 0.300 cm } BR;
				[ draw = WordBlueDarker, fill = WordBlueDarker, radius = 0.600 cm ] box {\color{white} \Large \sf{H}$^{ \otimes m }$} BR;
				hspace { 0.300 cm } BRA;
				[ draw = WordBlueDarker, fill = WordBlueDarker, radius = 0.600 cm ] box {\color{white} \Large \sf{H}$^{ \otimes m }$} BRA;
				[ name = Expansion, WordBlueVeryLight, line width = 0.500 mm, label = { [ label distance = 0.200 cm ] north: Expansion } ]
				barrier ( - ) ;
				cnot BR | BRA;
				[ name = Decryption, WordBlueVeryLight, line width = 0.500 mm, label = { [ label distance = 0.200 cm ] north: Decryption } ]
				barrier ( - ) ;
				[ line width = .250 mm, draw = white, fill = black, radius = 0.6 cm ] measure BR;
				hspace {0.500 cm} BR;
				output {$\ket{ \mathbf{ s } }$} BR;
				\node
				[
				bob,
				shirt = GreenLighter2,
				scale = 1.500,
				anchor = center,
				left = 0.250 cm of BOB,
				label = { [ label distance = 0.000 cm ] west: \textbf{Bob} }
				]
				(Bob) { };
				\node [ anchor = center, below = 2.500 cm of Bob ] (PhantomNode) { };
				\begin{scope} [ on background layer ]
					\draw
					[ MagentaDark, -, >=stealth, line width = 0.750 mm, decoration = coil, decorate ]
					( $ (IBREGA.east) + ( 0.500 mm, 0.000 mm ) $ ) node [ circle, fill, minimum size = 1.500 mm ] () {} -- ( $ (IBREG.east) + ( 0.500 mm, 0.000 mm ) $ ) node [ circle, fill, minimum size = 1.500 mm ] () {};
				\end{scope}
				\node [ below = 2.500 cm ] at (Input) { $\ket{ \psi_{ 1 } }$ };
				\node [ below = 2.500 cm ] at (Expansion) { $\ket{ \psi_{ 2 } }$ };
				\node [ below = 2.500 cm ] at (Decryption) { $\ket{ \psi_{ f } }$ };
			\end{yquant}
		\end{tikzpicture}
		\caption{This figure shows the quantum circuit used locally by Bob during the secret decryption phase to extract Alice's secret $\mathbf{ s }$.}
		\label{fig: $2$PSQDC Protocol Part 2}
	\end{figure}
\end{tcolorbox}

\begin{align}
	\label{eq: $2$PSQDC Part 2 Expansion 1}
	\ket{ \psi_{ 2 } }
	&=
	\frac { 1 } { \sqrt{ 2^{ m } } }
	\sum_{ \mathbf{ x } \in \mathbb{ B }^{ m } }
	\
	( - 1 )^{ \mathbf{ s } \cdot \mathbf{ x } }
	\
	H^{ \otimes m }
	\ket{ \mathbf{ x } }_{ BR_{ A } }
	\
	H^{ \otimes m }
	\ket{ \mathbf{ x } }_{ BR }
	\nonumber
	\\
	&\hspace{ -0.125 cm }
	\overset { ( \ref{eq: Hadamard m-Fold Ket x} ) } { = }
	\frac { 1 } { \sqrt{ 2^{ m } } }
	\sum_{ \mathbf{ x } \in \mathbb{ B }^{ m } }
	\
	( - 1 )^{ \mathbf{ s } \cdot \mathbf{ x } }
	\
	\left(
	\frac { 1 } { \sqrt{ 2^{ m } } }
	\sum_{ \mathbf{ a } \in \mathbb{ B }^{ m } }
	( - 1 )^{ \mathbf{ a } \cdot \mathbf{ x } }
	\ket{ \mathbf{ a } }_{ BR_{ A } }
	\right)
	\
	\left(
	\frac { 1 } { \sqrt{ 2^{ m } } }
	\sum_{ \mathbf{ b } \in  \mathbb{ B }^{ m } }
	( - 1 )^{ \mathbf{ b } \cdot \mathbf{ x } }
	\ket{ \mathbf{ b } }_{ BR }
	\right)
	\nonumber
	\\
	&=
	\frac { 1 } { 2^{ m } \sqrt{ 2^{ m } } }
	\sum_{ \mathbf{ a } \in \mathbb{ B }^{ m } }
	\sum_{ \mathbf{ b } \in \mathbb{ B }^{ m } }
	\sum_{ \mathbf{ x } \in \mathbb{ B }^{ m } }
	\
	( - 1 )^{ ( \mathbf{ s } \oplus \mathbf{ a } \oplus \mathbf{ b } ) \cdot \mathbf{ x } }
	\
	\ket{ \mathbf{ a } }_{ BR_{ A } }
	\
	\ket{ \mathbf{ b } }_{ BR }
	\ .
\end{align}

The above expression \eqref{eq: $2$PSQDC Part 2 Expansion 1} can be simplified via the use of the Characteristic Inner Product property \eqref{eq: The Characteristic Inner Product Modulo $2$ Property} of the inner product modulo $2$, which asserts that if

\begin{align}
	\label{eq: $2$ Player Fundamental Correlation Property}
	\mathbf{ a }
	\oplus
	\mathbf{ b }
	\oplus
	\mathbf{ s }
	=
	\mathbf{ 0 }
	\Leftrightarrow
	\mathbf{ a }
	\oplus
	\mathbf{ b }
	=
	\mathbf{ s }
	\ ,
\end{align}

the sum $\sum_{ \mathbf{ x } \in \mathbb{ B }^{ m } } ( - 1 )^{ ( \mathbf{ s } \oplus \mathbf{ a } \oplus \mathbf{ b } ) \cdot \mathbf{ x } } \ket{ \mathbf{ a } }_{ BR_{ A } } \ket{ \mathbf{ b } }_{ BR }$ is equal to $2^{ m }$ $\ket{ \mathbf{ a } }_{ BR_{ A } }$ $\ket{ \mathbf{ b } }_{ BR }$, whereas if $\mathbf{ a } \oplus \mathbf{ b } \oplus \mathbf{ s } \neq \mathbf{ 0 }$, the sum reduces to $0$. This allows the rewriting of $\ket{ \psi_{ 2 } }$ as

\begin{align}
	\label{eq: $2$PSQDC Part 2 Expansion 2}
	\ket{ \psi_{ 2 } }
	=
	\frac { 1 } { \sqrt{ 2^{ m } } }
	\sum_{ \mathbf{ a } \in \mathbb{ B }^{ m } }
	\sum_{ \mathbf{ b } \in \mathbb{ B }^{ m } }
	\
	\ket{ \mathbf{ a } }_{ BR_{ A } }
	\
	\ket{ \mathbf{ b } }_{ BR }
	\ ,
	\quad \text{where} \quad
	\mathbf{ a } \oplus \mathbf{ b } = \mathbf{ s }
	\ .
\end{align}

By recalling \eqref{eq: Secret Bit Vector}, and utilizing the bit form of $\mathbf{ a } = a_{ m - 1 } \dots a_{ 0 }$ and $\mathbf{ b } = b_{ m - 1 } \dots b_{ 0 }$, we may write the explicit bitwise version of \eqref{eq: $2$ Player Fundamental Correlation Property}:

\begin{align}
	\label{eq: $2$ Player Bitwise Fundamental Correlation Property}
	\left
	\{
	\
	\begin{aligned}
		a_{ m - 1 }
		\oplus
		b_{ m - 1 }
		&=
		s_{ m - 1 }
		\\
		\dots
		\\
		a_{ 0 }
		\oplus
		b_{ 0 }
		&=
		s_{ 0 }
	\end{aligned}
	\
	\right
	\}
	\ .
\end{align}

Equations \eqref{eq: $2$ Player Fundamental Correlation Property} and \eqref{eq: $2$ Player Bitwise Fundamental Correlation Property} give the mathematical expression of the correlation of the contents of the registers $BR$ and $BR_{ A }$. This is a result of the entanglement between Alice and Bob's registers in the initial state of the quantum circuit of Figure \ref{fig: $2$PSQDC Protocol Part 1}. The practical significance of this fact is that the contents of the two registers do not vary independently of each other, as they must always obey \eqref{eq: $2$ Player Fundamental Correlation Property} and \eqref{eq: $2$ Player Bitwise Fundamental Correlation Property}, which from now on we shall refer to as the $2$ player \textbf{Hadamard Entanglement Property} and the $2$ player \textbf{Bitwise Hadamard Entanglement Property}, respectively.

Let us now recall that a CNOT (controlled-NOT) gate acts by negating the second qubit, called target qubit, if and only if the first qubit, called control qubit, is in state $\ket{ 1 }$. One way to express its operation is by writing

\begin{align}
	\label{eq: CNOT Gate}
	{ \rm CNOT }
	\ket{ q_{ c } }
	\ket{ q_{ t } }
	=
	\ket{ q_{ c } }
	\ket{ q_{ t } \oplus q_{ c } }
	\ ,
\end{align}

where $q_{ c }$ and $q_{ t }$ designate the control and target qubits respectively.

Having established notation, we now proceed to explain in detail Bob's actions to decipher Alice's secret. Bob applies $m$ CNOT gates to his registers $BR_{ A }$ and $BR$. Each of the $m$ qubits in $BR_{ A }$ will serve as a control qubit targeting the corresponding qubit in $BR$. This process drives the circuit to its final state $\ket{ \psi_{ f } }$, and its formal mathematical description is given below.

\begin{align}
	\label{eq: $2$PSQDC Part 2 Decryption 1}
	&{ \rm CNOT }
	\ket{ a_{ m - 1 } }
	\ket{ b_{ m - 1 } }
	\dots
	{ \rm CNOT }
	\ket{ a_{ 0 } }
	\ket{ b_{ 0 } }
	\nonumber
	\\
	&=
	\ket{ a_{ m - 1 } }
	\ket{ a_{ m - 1 } \oplus b_{ m - 1 } }
	\dots
	\ket{ a_{ 0 } }
	\ket{ a_{ 0 } \oplus b_{ 0 } }
	\nonumber \\
	&\overset { \eqref{eq: $2$ Player Bitwise Fundamental Correlation Property} } { = }
	\ket{ \mathbf{ a } }_{ 1 }
	\ket{ \mathbf{ s } }_{ 0 }
	\ .
\end{align}

Hence, in view of equations \eqref{eq: $2$PSQDC Part 2 Expansion 2} and \eqref{eq: $2$PSQDC Part 2 Decryption 1}, the final state $\ket{ \psi_{ f } }$ can be expressed as

\begin{align}
	\label{eq: $2$PSQDC Part 2 Decryption 2}
	\ket{ \psi_{ 3 } }
	=
	\frac { 1 } { \sqrt{ 2^{ m } } }
	\sum_{ \mathbf{ a } \in \mathbb{ B }^{ m } }
	\ket{ \mathbf{ a } }_{ BR_{ A } }
	\ket{ \mathbf{ s } }_{ BR }
	\ .
\end{align}

Now, Bob measures (in the computational basis) the contents of the register $BR$ and obtains the secret bit vector $\mathbf{ s }$.

\section{The $3$PSQDC protocol} \label{sec: The $3$PSQDC Protocol}

The $3$ player protocol for quantum secure direct communication, $3$PSQDC from now on, is a generalization of the $2$PSQDC scheme, where three spatially separated players interact in order to enable two of them to communicate with the third player directly and securely using only the quantum channel. Many real-life situations involve more than two agents. Therefore, it is advantageous to possess algorithms and techniques facilitating secure communication and information exchange among an arbitrary number of players.

The $3$PSQDC protocol evolves as a game among Alice, Bob and Charlie. To make the game more interesting this time, we assume that Bob and Charlie, being loyal agents of Alice, each have come up with some information, which is incomplete by itself. Only by combining the two pieces of information can the complete secret be revealed. Therefore, Bob and Charlie must send their information to Alice, so that she may uncover the secret. A similar setting in which an arbitrary number of agents send information to Alice in order for her to compose the complete secret, was analyzed in \cite{Ampatzis2023}. The major difference compared to the present work, is that the protocol in \cite{Ampatzis2023} doesn't involve direct quantum communication because at the final stage the agents communicate to Alice the information necessary to unlock the secret through the classical channel. Eve, as usual, undertakes the role of the adversary aiming to sabotage the protocol and steal the secret.

\subsection{Entanglement distribution phase} \label{subsec: $3$PSQDC Entanglement Distribution Phase}

We may conceptually divide the $3$PSQDC protocol in phases. The first phase is the entanglement distribution phase, during which Alice, or a third trusted source, prepares $\ket{ GHZ_{ 3 } }$ triplets. For the execution of the protocol, it is immaterial whether it is Alice or another trusted source that undertakes this task. Using a modern quantum computer, or some equivalent apparatus, it is easy to produce entangled triplets. These triplets are subsequently shared among Alice, Bob, and Charlie, according to the $3$-Uniform Distribution Scheme explained in Definition \ref{def: Uniform Entanglement Distribution Scheme}, i.e., $r = 3$ in this case.

The technicalities regarding the additional GHZ triplets that serve for the eavesdropping detection and entanglement validation tests are omitted because we strive to ease the presentation of the $3$PSQDC protocol. The interested reader is referred to the suggested bibliography mentioned in subsection \ref{subsec: The Role of the Classical Channel & Defensive Measures}. The situation before Alice distributes Bob and Charlie's sequences is visualized in Figure \ref{fig:Alice Creates $m$ GHZ_{ 3 } Triplets}. At the end of the distribution phase, Alice, Bob and Charlie have $m$ qubits in their local quantum registers each. All three registers are correlated because Alice, Bob and Charlie's corresponding qubits are entangled in the $\ket{ GHZ_{ 3 } }$ state, and the whole setup is shown in Figure \ref{fig: Alice Shares $m$ GHZ_{ 3 } Triplets with Bob and Charlie}.

\begin{tcolorbox}
	[
		grow to left by = 0.000 cm,
		grow to right by = 0.000 cm,
		colback = WordTurquoiseLighter80!12,			
		enhanced jigsaw,								
		sharp corners,
		toprule = 1.000 pt,
		bottomrule = 1.000 pt,
		leftrule = 0.100 pt,
		rightrule = 0.100 pt,
		sharp corners,
		center title,
		fonttitle = \bfseries
	]
	\begin{figure}[H]
		\begin{minipage} [ b ] { 0.475 \textwidth }
			\centering
			\begin{tikzpicture} [ scale = 0.500 ]
				\node
					[
						shade, top color = MagentaLighter, bottom color = black, rectangle, text width = 5.000 cm, align = center
					]
					(Label)
					{ \color{white} $\ket{ GHZ_{ 3 } }$ Triplets Preparation };
				\node
					[
						alice,
						shirt = RedPurple,
						scale = 1.500,
						anchor = center,
						below = 1.000 cm of Label,
						label = { [ label distance = 0.000 cm ] north: \textbf{Alice} }
					]
					(Alice) { };
				\node
					[
						bob,
						scale = 1.500,
						shirt = GreenLighter2,
						anchor = center,
						left = 2.500 cm of Alice,
						label = { [ label distance = 0.000 cm ] north: \textbf{Bob} }
					]
					(Bob) { };
				\node
					[
						charlie,
						scale = 1.500,
						shirt = WordBlueVeryLight,
						anchor = center,
						right = 2.500 cm of Alice,
						label = { [ label distance = 0.000 cm ] north: \textbf{Charlie} }
					]
					(Charlie) { };
				\node
					[
						circle, shade, outer color = RedPurple!50, inner color = white, minimum size = 5.000 mm, below = 1.000 cm of Alice,
						label = { [ label distance = 0.000 cm ] north: $\ket{ GHZ_{ 3 } }_{ m - 1 }$ }
					]
					(GHZ3-A-m-1) { };
				\node
					[
						circle, shade, outer color = GreenLighter2!50, inner color = white, minimum size = 5.000 mm, left = 7.000 mm of GHZ3-A-m-1,
					]
					(GHZ3-B-m-1) { };
				\node
					[
						circle, shade, outer color = WordBlueVeryLight!50, inner color = white, minimum size = 5.000 mm, right = 7.000 mm of GHZ3-A-m-1,
					]
					(GHZ3-C-m-1) { };
				\scoped [ on background layer ]
					\draw
						[ MagentaDark, -, >=stealth, line width = 0.750 mm, decoration = coil, decorate ]
						(GHZ3-B-m-1) -- (GHZ3-A-m-1) -- (GHZ3-C-m-1);
				\node
					[
						minimum size = 5.000 mm, below = 0.250 cm of GHZ3-A-m-1,
					]
					(VDotsA)
					{ \vdots };
				\node
					[
						minimum size = 5.000 mm, left = 7.000 mm of VDotsA,
					]
					(VDotsB)
					{ \vdots };
				\node
					[
						minimum size = 5.000 mm, right = 7.000 mm of VDotsA,
					]
					(VDotsC)
					{ \vdots };
				\node
					[
						circle, shade, outer color = RedPurple!50, inner color = white, minimum size = 5.000 mm, below = 1.000 cm of VDotsA,
						label = { [ label distance = 0.000 cm ] north: $\ket{ GHZ_{ 3 } }_{ 0 }$ }
					]
					(GHZ3-A-0) { };
				\node
					[
						circle, shade, outer color = GreenLighter2!50, inner color = white, minimum size = 5.000 mm, left = 7.000 mm of GHZ3-A-0,
					]
					(GHZ3-B-0) { };
				\node
					[
						circle, shade, outer color = WordBlueVeryLight!50, inner color = white, minimum size = 5.000 mm, right = 7.000 mm of GHZ3-A-0,
					]
					(GHZ3-C-0) { };
				\scoped [ on background layer ]
					\draw
						[ MagentaDark, -, >=stealth, line width = 0.750 mm, decoration = coil, decorate ]
						(GHZ3-B-0) -- (GHZ3-A-0) -- (GHZ3-C-0);
				\draw
					[ WordBlueVeryLight, <->, >=stealth, line width = 2.500 mm ]
					( $ (Alice.west) - ( 3.000 mm, 0.000 mm ) $ ) -- ( $ (Bob.east) + ( 3.000 mm, 0.000 mm ) $ )
					node [ above = 0.250, midway ] { };
				\draw
					[ WordBlueVeryLight, <->, >=stealth, line width = 2.500 mm, shade, left color = WordBlueVeryLight, right color = white ]
					(Alice) -- (Charlie) node [ above, midway, text = black ] { };
				\node [ anchor = center, below = 0.000 mm of GHZ3-A-0 ] (PhantomNode1) { };
			\end{tikzpicture}
			\caption{Alice creates the $m$ $\ket{ GHZ_{ 3 } }$ triplets that she will share with Bob and Charlie, who are in different spatial locations.}
			\label{fig:Alice Creates $m$ GHZ_{ 3 } Triplets}
		\end{minipage}
		\hfill
		\begin{minipage} [ b ] { 0.475 \textwidth }
			\centering
			\begin{tikzpicture} [ scale = 0.500 ]
				\node
					[
						shade, top color = MagentaLighter, bottom color = black, rectangle, text width = 5.000 cm, align = center
					]
					(Label)
					{ \color{white} $\ket{ GHZ_{ 3 } }$ Triplet Distribution };
				\node
					[
						alice,
						shirt = RedPurple,
						scale = 1.500,
						anchor = center,
						below = 1.000 cm of Label,
						label = { [ label distance = 0.000 cm ] north: \textbf{Alice} }
					]
					(Alice) { };
				\node
					[
						bob,
						scale = 1.500,
						shirt = GreenLighter2,
						anchor = center,
						left = 2.500 cm of Alice,
						label = { [ label distance = 0.000 cm ] north: \textbf{Bob} }
					]
					(Bob) { };
				\node
					[
						charlie,
						scale = 1.500,
						shirt = WordBlueVeryLight,
						anchor = center,
						right = 2.500 cm of Alice,
						label = { [ label distance = 0.000 cm ] north: \textbf{Charlie} }
					]
					(Charlie) { };
				\node
					[
						circle, shade, outer color = RedPurple!50, inner color = white, minimum size = 5.000 mm, below = 1.000 cm of Alice,
						label = { [ label distance = 0.000 cm ] north: $\ket{ GHZ_{ 3 } }_{ m - 1 }$ }
					]
					(GHZ3-A-m-1) { };
				\node
					[
						circle, shade, outer color = GreenLighter2!50, inner color = white, minimum size = 5.000 mm, below = 1.000 cm of Bob,
					]
					(GHZ3-B-m-1) { };
				\node
					[
						circle, shade, outer color = WordBlueVeryLight!50, inner color = white, minimum size = 5.000 mm, below = 1.000 cm of Charlie,
					]
					(GHZ3-C-m-1) { };
				\scoped [ on background layer ]
					\draw
					[ MagentaDark, -, >=stealth, line width = 0.750 mm, decoration = coil, decorate ]
					(GHZ3-B-m-1) -- (GHZ3-A-m-1) -- (GHZ3-C-m-1);
				\node
					[
						minimum size = 5 mm, below = 0.250 cm of GHZ3-A-m-1,
					]
					(VDotsA)
					{ \vdots };
				\node
					[
						minimum size = 5 mm, below = 0.250 cm of GHZ3-B-m-1,
					]
					(VDotsB)
					{ \vdots };
				\node
					[
						minimum size = 5 mm, below = 0.250 cm of GHZ3-C-m-1,
					]
					(VDotsC)
					{ \vdots };
				\node
					[
						circle, shade, outer color = RedPurple!50, inner color = white, minimum size = 5 mm, below = 1.00 cm of VDotsA,
						label = { [ label distance = 0.00 cm ] north: $\ket{ GHZ_{ 3 } }_{ 0 }$ }
					]
					(GHZ3-A-0) { };
				\node
					[
						circle, shade, outer color = GreenLighter2!50, inner color = white, minimum size = 5 mm, below = 1.00 cm of VDotsB,
					]
					(GHZ3-B-0) { };
				\node
					[
						circle, shade, outer color = WordBlueVeryLight!50, inner color = white, minimum size = 5 mm, below = 1.00 cm of VDotsC,
					]
					(GHZ3-C-0) { };
				\scoped [ on background layer ]
					\draw
						[ MagentaDark, -, >=stealth, line width = 0.750 mm, decoration = coil, decorate ]
						(GHZ3-B-0) -- (GHZ3-A-0) -- (GHZ3-C-0);
				\draw
					[ WordBlueVeryLight, <->, >=stealth, line width = 2.500 mm ]
					( $ (Alice.west) - ( 3.000 mm, 0.000 mm ) $ ) -- ( $ (Bob.east) + ( 3.000 mm, 0.000 mm ) $ )
					node [ above = 0.250, midway ] { };
				\draw
					[ WordBlueVeryLight, <->, >=stealth, line width = 2.500 mm, shade, left color = WordBlueVeryLight, right color = white ]
					(Alice) -- (Charlie) node [ above, midway, text = black ] { };
				\node [ anchor = center, below = 0.000 mm of GHZ3-A-0 ] (PhantomNode1) { };
			\end{tikzpicture}
			\caption{Alice shares the $m$ $\ket{ GHZ_{ 3 } }$ triplets with Bob and Charlie by keeping the first and sending the second and third qubits to Bob and Charlie.}
			\label{fig: Alice Shares $m$ GHZ_{ 3 } Triplets with Bob and Charlie}
		\end{minipage}
	\end{figure}
\end{tcolorbox}


At the end of this phase, Alice, Bob, and Charlie execute the eavesdropping detection and entanglement validation tests utilizing the classical channel. If these tests are successful, they proceed to the secret embedding phase. If not, they must assume that Eve has disrupted their communication and start all over again, after implementing corrective measures.

\subsection{Secret embedding phase} \label{subsec: $3$PSQDC Secret Embedding Phase}

The idea behind this generalization is that the two agents Bob and Charlie have in their possession part of a secret that they must transmit to Alice in order for her to obtain the complete secret $\mathbf{ s }$. To succeed in this task, the three players use their local circuits outlined in Figure \ref{fig: $3$PSQDC Protocol Part 1}. Despite their spatial separation, the correlations among their registers, due to entanglement, effectively create a single distributed system.

In Figure \ref{fig: $3$PSQDC Protocol Part 1}, $AR$, $BR$ and $CR$ stand for Alice, Bob and Charlie's entangled registers, respectively, while $BQ$ and $CQ$ denote Bob and Charlie's qubits, both initialized in state $\ket{ - }$. To avoid any confusion, we use subscripts $A$, $B$ and $C$ to distinguish among Alice, Bob and Charlie's qubits and registers.

The partial secrets that Bob and Charlie intend to convey to Alice are encoded in the secret bit vectors $\mathbf{ s }_{ B }$ and  $\mathbf{ s }_{ C }$.
Alice must combine both of them via bitwise addition in order to uncover the complete secret $\mathbf{ s }$:

\begin{align}
	\label{eq: Alice Bob and Charlie's Secret Bit Vectors}
	\mathbf{ s }
	=
	\mathbf{ s }_{ B }
	\oplus
	\mathbf{ s }_{ B }
	\ .
\end{align}

\begin{tcolorbox}
	[
		enhanced,
		breakable,
		grow to left by = 0.000 cm,
		grow to right by = 0.000 cm,
		colback = MagentaLighter!07,			
		enhanced jigsaw,						
		sharp corners,
		toprule = 1.000 pt,
		bottomrule = 1.000 pt,
		leftrule = 0.100 pt,
		rightrule = 0.100 pt,
		sharp corners,
		center title,
		fonttitle = \bfseries
	]
	\begin{figure}[H]
		\centering
		\begin{tikzpicture} [ scale = 1.000 ]
			\begin{yquant}
				nobit AUX_C_0;
				nobit AUX_C_1;
				[ name = CREG ] qubits { $CR$ \hspace{ 0.000 cm } } CR;
				nobit AUX_C_2;
				qubit { $CQ$: \hspace{ 0.000 cm } $\ket{ - }$ } CQ;
				nobit AUX_C_3;
				nobit AUX_C_4;
				nobit AUX_C_5;
				nobit AUX_B_0;
				nobit AUX_B_1;
				nobit AUX_B_2;
				[ name = BREG ] qubits { $BR$ \hspace{ 0.000 cm } } BR;
				nobit AUX_B_4;
				qubit { $BQ$: \hspace{ 0.000 cm } $\ket{ - }$ } BQ;
				nobit AUX_B_5;
				nobit AUX_B_6;
				nobit AUX_A_0;
				nobit AUX_A_1;
				nobit AUX_A_3;
				nobit AUX_A_4;
				[ name = AREG ] qubits { $AR$ \hspace{ 0.000 cm } } AR;
				nobit AUX_A_5;
				nobit AUX_A_6;
				[ name = InitialState, WordBlueVeryLight, line width = 0.500 mm, label = { [ label distance = 0.200 cm ] north: Initial State } ]
				barrier ( - ) ;
				hspace { 0.500 cm } CR;
				[ draw = RedPurple, fill = RedPurple, radius = 0.700 cm ] box {\color{white} \Large \sf{U}$_{ C }$} (CR - CQ);
				hspace { 0.500 cm } BR;
				[ draw = RedPurple, fill = RedPurple, radius = 0.700 cm ] box {\color{white} \Large \sf{U}$_{ B }$} (BR - BQ);
				[ name = SecretEmbedding, WordBlueVeryLight, line width = 0.500 mm, label = { [ label distance = 0.100 cm ] north: Secret Embedding } ]
				barrier ( - ) ;
				output { } CR;
				output { } BR;
				output { } AR;
				\node
					[
						charlie,
						shirt = WordBlueVeryLight,
						scale = 1.500,
						anchor = center,
						left = 0.250 cm of CREG,
						label = { [ label distance = 0.000 cm ] north: \textbf{Charlie} }
					]
					(Charlie) { };
				\node
					[
						bob,
						shirt = GreenLighter2,
						scale = 1.500,
						anchor = center,
						left = 0.250 cm of BREG,
						label = { [ label distance = 0.000 cm ] north: \textbf{Bob} }
					]
					(Bob) { };
				\node
					[
						alice,
						shirt = RedPurple,
						scale = 1.500,
						anchor = center,
						left = 0.250 cm of AREG,
						label = { [ label distance = 0.000 cm ] north: \textbf{Alice} }
					]
					(Alice) { };
				\begin{scope} [ on background layer ]
					\node
						[ above right = 0.375 cm and 1.500 cm of Alice, rectangle, fill = WordAquaLighter60, text width = 6.000 cm, align = center, minimum height = 10.000 mm ] { \bf Spatially Separated };
					\node
						[ above right = 0.500 cm and 1.500 cm of Bob, rectangle, fill = WordAquaLighter60, text width = 6.000 cm, align = center, minimum height = 10.000 mm ] { \bf Spatially Separated };
					\draw
						[ MagentaDark, -, >=stealth, line width = 0.750 mm, decoration = coil, decorate ]
						( $ (AREG.east) + ( 0.500 mm, 0.000 mm ) $ ) node [ circle, fill, minimum size = 1.500 mm ] () {} -- ( $ (BREG.east) + ( 0.500 mm, 0.000 mm ) $ ) node [ circle, fill, minimum size = 1.500 mm ] () {} -- ( $ (CREG.east) + ( 0.500 mm, 0.000 mm ) $ ) node [ circle, fill, minimum size = 1.500 mm ] () {};
				\end{scope}
				\node [ below = 4.750 cm ] at (InitialState) { $\ket{ \psi_{ 0 } }$ };
				\node [ below = 4.750 cm ] at (SecretEmbedding) { $\ket{ \psi_{ 1 } }$ };
				\node [ anchor = center, below = 0.500 cm of Alice ] (PhantomNode) { };
			\end{yquant}
		\end{tikzpicture}
		\caption{This figure presents an outline of the quantum circuits employed by Alice, Bob and Charlie during the secret embedding phase. This distributed system evolves from the initial state $\ket{ \psi_{ 0 } }$ to the state $\ket{ \psi_{ 1 } }$.}
		\label{fig: $3$PSQDC Protocol Part 1}
	\end{figure}
\end{tcolorbox}

Bob and Charlie encode their secrets into the state of the compound system by using the unitary transforms $U_{ B }$ and $U_{ C }$ on their registers, where

\begin{align}
	\label{eq: Bob and Charlie's Explicit Unitary Transforms}
	\left
	\{
	\
	\begin{aligned}
		U_{ B }
		\colon
		\ket{ - }_{ B }
		\
		\ket{ \mathbf{ x } }_{ B }
		&\rightarrow
		( - 1 )^{ \mathbf{ s }_{ B } \cdot \mathbf{ x } }
		\
		\ket{ - }_{ B }
		\
		\ket{ \mathbf{ x } }_{ B }
		\\
		&\text{ and }
		\\
		U_{ C }
		\colon
		\ket{ - }_{ C }
		\
		\ket{ \mathbf{ x } }_{ C }
		&\rightarrow
		( - 1 )^{ \mathbf{ s }_{ C } \cdot \mathbf{ x } }
		\
		\ket{ - }_{ C }
		\
		\ket{ \mathbf{ x } }_{ C }
	\end{aligned}
	\
	\right
	\}
	\ .
\end{align}

Using \eqref{eq: m-Fold GHZ 3 States} we can express the initial state $\ket{ \psi_{ 0 } }$ of the circuit shown in Figure \ref{fig: $3$PSQDC Protocol Part 1} as

\begin{align}
	\label{eq: $3$PSQDC Part 1 Initial State}
	\ket{ \psi_{ 0 } }
	=
	\frac { 1 } { \sqrt{ 2^{ m } } }
	\sum_{ \mathbf{ x } \in \mathbb{ B }^{ m } }
	\
	\ket{ \mathbf{ x } }_{ A }
	\
	\ket{ - }_{ B }
	\
	\ket{ \mathbf{ x } }_{ B }
	\
	\ket{ - }_{ C }
	\
	\ket{ \mathbf{ x } }_{ C }
	\ .
\end{align}

The application of the unitary transforms \eqref{eq: Bob and Charlie's Explicit Unitary Transforms} sends the system to the next state $\ket{ \psi_{ 1 } }$:

\begin{align}
	\label{eq: $3$PSQDC Part 1 Secret Embedding State}
	\ket{ \psi_{ 1 } }
	&\overset { \eqref{eq: Bob and Charlie's Explicit Unitary Transforms} } { = }
	\frac { 1 } { \sqrt{ 2^{ m } } }
	\sum_{ \mathbf{ x } \in \mathbb{ B }^{ m } }
	\
	\ket{ \mathbf{ x } }_{ A }
	\
	( - 1 )^{ \mathbf{ s }_{ B } \cdot \mathbf{ x } }
	\
	\ket{ - }_{ B }
	\
	\ket{ \mathbf{ x } }_{ B }
	\
	( - 1 )^{ \mathbf{ s }_{ C } \cdot \mathbf{ x } }
	\
	\ket{ - }_{ C }
	\
	\ket{ \mathbf{ x } }_{ C }
	\nonumber
	\\
	&\overset { \eqref{eq: Alice Bob and Charlie's Secret Bit Vectors} } { = }
	\frac { 1 } { \sqrt{ 2^{ m } } }
	\sum_{ \mathbf{ x } \in \mathbb{ B }^{ m } }
	\
	( - 1 )^{ \mathbf{ s } \cdot \mathbf{ x } }
	\
	\ket{ \mathbf{ x } }_{ A }
	\
	\ket{ - }_{ B }
	\
	\ket{ \mathbf{ x } }_{ B }
	\
	\ket{ - }_{ C }
	\
	\ket{ \mathbf{ x } }_{ C }
	\ .
\end{align}

At the end of this phase, Bob sends to Alice the $m$ qubits in his register through the quantum channel. Alice organizes these qubits in an additional register denoted by $AR_{ B }$. Similarly, Alice receives from Charlie his $m$ qubits and places them into a third register, designated by $AR_{ C }$. Before Alice implements the final phase of the protocol, it is crucial that she and her agents conduct again the eavesdropping detection and entanglement validation tests. If both tests are successful, then Alice may proceed to decode the secret information. If not, then they should assume the worst case scenario, i.e., that Eve has compromised their communication and start all from scratch, after enforcing enhanced security.

\subsection{Secret decryption phase} \label{subsec: $3$PSQDC Secret Decryption Phase}

\begin{tcolorbox}
	[
		enhanced,
		breakable,
		grow to left by = 0.000 cm,
		grow to right by = 0.000 cm,
		colback = MagentaLighter!07,			
		enhanced jigsaw,						
		sharp corners,
		toprule = 1.000 pt,
		bottomrule = 1.000 pt,
		leftrule = 0.100 pt,
		rightrule = 0.100 pt,
		sharp corners,
		center title,
		fonttitle = \bfseries
	]
	\begin{figure}[H]
		\centering
		\begin{tikzpicture}[ scale = 1.000 ]
			\begin{yquant}
				nobit AUX_C_0;
				[ name = AREGC ] qubits  { $AR_{ C }$ \hspace{ 0.000 cm } } ARC;
				nobit AUX_B_0;
				[ name = AREGB ] qubits  { $AR_{ B }$ \hspace{ 0.000 cm } } ARB;
				nobit AUX_A_0;
				[ name = AREG ] qubits { $AR$ \hspace{ 0.000 cm } } AR;
				nobit AUX_A_1;
				[ name = Input, WordBlueVeryLight, line width = 0.500 mm, label = { [ label distance = 0.200 cm ] north: Input } ]
				barrier ( - ) ;
				hspace { 0.300 cm } -;
				[ draw = WordBlueDarker, fill = WordBlueDarker, radius = 0.600 cm ] box {\color{white} \Large \sf{H}$^{ \otimes m }$} ARC;
				[ draw = WordBlueDarker, fill = WordBlueDarker, radius = 0.600 cm ] box {\color{white} \Large \sf{H}$^{ \otimes m }$} ARB;
				[ draw = WordBlueDarker, fill = WordBlueDarker, radius = 0.600 cm ] box {\color{white} \Large \sf{H}$^{ \otimes m }$} AR;
				[ name = Expansion, WordBlueVeryLight, line width = 0.500 mm, label = { [ label distance = 0.200 cm ] north: Expansion } ]
				barrier ( - ) ;
				cnot ARB | AR;
				hspace { 0.300 cm } -;
				cnot ARC | ARB;
				[ name = Decryption, WordBlueVeryLight, line width = 0.500 mm, label = { [ label distance = 0.200 cm ] north: Decryption } ]
				barrier ( - ) ;
				[ line width = .250 mm, draw = white, fill = black, radius = 0.6 cm ] measure ARC;
				output {$\ket{ \mathbf{ s } }$} ARC;
				\node
					[
						alice,
						shirt = RedPurple,
						scale = 1.500,
						anchor = center,
						left = 0.250 cm of AREGB,
						label = { [ label distance = 0.000 cm ] north: \textbf{Alice} }
					]
					(Alice) { };
				\begin{scope} [ on background layer ]
					\draw
						[ MagentaDark, -, >=stealth, line width = 0.750 mm, decoration = coil, decorate ]
						( $ (AREG.east) + ( 0.500 mm, 0.000 mm ) $ ) node [ circle, fill, minimum size = 1.500 mm ] () {} -- ( $ (AREGB.east) + ( 0.500 mm, 0.000 mm ) $ ) node [ circle, fill, minimum size = 1.500 mm ] () {} -- ( $ (AREGC.east) + ( 0.500 mm, 0.000 mm ) $ ) node [ circle, fill, minimum size = 1.500 mm ] () {};
				\end{scope}
				\node [ below = 2.750 cm ] at (Input) { $\ket{ \psi_{ 1 } }$ };
				\node [ below = 2.750 cm ] at (Expansion) { $\ket{ \psi_{ 2 } }$ };
				\node [ below = 2.750 cm ] at (Decryption) { $\ket{ \psi_{ f } }$ };
				\node [ anchor = center, below = 2.500 cm of Alice ] (PhantomNode) { };
			\end{yquant}
		\end{tikzpicture}
		\caption{This figure shows the quantum circuit that Alice uses locally during the decryption phase to extract the complete secret $\mathbf{ s }$.}
		\label{fig: $3$PSQDC Protocol Part 2}
	\end{figure}
\end{tcolorbox}

During this phase, Alice uses the quantum circuit outlined in Figure \ref{fig: $3$PSQDC Protocol Part 2} to decrypt the complete secret bit vector $\mathbf{ s }$. In addition to the quantum register $AR$, shown in the circuit of Figure \ref{fig: $3$PSQDC Protocol Part 1}, Alice now employs two additional registers $AR_{ B }$ and $AR_{ C }$. All three registers are entangled and, by using all of them, Alice will uncover the secret bit vector $\mathbf{ s }$. To avoid any confusion, in the formulae below we employ subscripts $AR_{ C }$, $AR_{ B }$, and $AR$ to designate registers $AR_{ C }$, $AR_{ B }$ and $AR$, respectively. Initially, the circuit depicted in Figure \ref{fig: $3$PSQDC Protocol Part 2} is in state $\ket{ \psi_{ 1 } }$. Alice applies to all her registers $m$-fold Hadamard transforms, driving the system to the next state $\ket{ \psi_{ 2 } }$.

\begin{align}
	\hspace{ -2.000 cm }
	\label{eq: $3$PSQDC Part 2 Expansion 1}
	\ket{ \psi_{ 2 } }
	&=
	\frac { 1 } { \sqrt{ 2^{ m } } }
	\sum_{ \mathbf{ x } \in \mathbb{ B }^{ m } }
	\
	( - 1 )^{ \mathbf{ s } \cdot \mathbf{ x } }
	\
	H^{ \otimes m }
	\ket{ \mathbf{ x } }_{ AR }
	\
	H^{ \otimes m }
	\ket{ \mathbf{ x } }_{ AR_{ B } }
	\
	H^{ \otimes m }
	\ket{ \mathbf{ x } }_{ AR_{ C } }
	\nonumber \\
	&\hspace{ -0.125 cm }
	\overset { ( \ref{eq: Hadamard m-Fold Ket x} ) } { = }
	\frac { 1 } { \sqrt{ 2^{ m } } }
	\sum_{ \mathbf{ x } \in \mathbb{ B }^{ m } }
	\
	( - 1 )^{ \mathbf{ s } \cdot \mathbf{ x } }
	\
	\left(
	\frac { 1 } { \sqrt{ 2^{ m } } }
	\sum_{ \mathbf{ a } \in \mathbb{ B }^{ m } }
	( - 1 )^{ \mathbf{ a } \cdot \mathbf{ x } }
	\ket{ \mathbf{ a } }_{ AR }
	\right)
	\
	\left(
	\frac { 1 } { \sqrt{ 2^{ m } } }
	\sum_{ \mathbf{ b } \in  \mathbb{ B }^{ m } }
	( - 1 )^{ \mathbf{ b } \cdot \mathbf{ x } }
	\ket{ \mathbf{ b } }_{ AR_{ B } }
	\right)
	\
	\left(
	\frac { 1 } { \sqrt{ 2^{ m } } }
	\sum_{ \mathbf{ c } \in  \mathbb{ B }^{ m } }
	( - 1 )^{ \mathbf{ c } \cdot \mathbf{ x } }
	\ket{ \mathbf{ c } }_{ AR_{ C } }
	\right)
	\nonumber \\
	&=
	\frac { 1 } { 2^{ 2 m } }
	\sum_{ \mathbf{ a } \in \mathbb{ B }^{ m } }
	\sum_{ \mathbf{ b } \in \mathbb{ B }^{ m } }
	\sum_{ \mathbf{ c } \in \mathbb{ B }^{ m } }
	\sum_{ \mathbf{ x } \in \mathbb{ B }^{ m } }
	\
	( - 1 )^{ ( \mathbf{ s } \oplus \mathbf{ a } \oplus \mathbf{ b } \oplus \mathbf{ c } ) \cdot \mathbf{ x } }
	\
	\ket{ \mathbf{ a } }_{ AR }
	\
	\ket{ \mathbf{ b } }_{ AR_{ B } }
	\
	\ket{ \mathbf{ c } }_{ AR_{ C } }
	\ .
\end{align}

Applying the Characteristic Inner Product property \eqref{eq: The Characteristic Inner Product Modulo $2$ Property}, which asserts that if

\begin{align}
	\label{eq: $3$ Player Fundamental Correlation Property}
	\mathbf{ a }
	\oplus
	\mathbf{ b }
	\oplus
	\mathbf{ c }
	\oplus
	\mathbf{ s }
	=
	\mathbf{ 0 }
	\Leftrightarrow
	\mathbf{ a }
	\oplus
	\mathbf{ b }
	\oplus
	\mathbf{ c }
	=
	\mathbf{ s }
	\ ,
\end{align}

the sum $\sum_{ \mathbf{ x } \in \mathbb{ B }^{ m } }$ $( - 1 )^{ ( \mathbf{ s } \oplus \mathbf{ a } \oplus \mathbf{ b } \oplus \mathbf{ c } ) \cdot \mathbf{ x } }$ $\ket{ \mathbf{ a } }_{ AR }$ $\ket{ \mathbf{ b } }_{ AR_{ B } }$ $\ket{ \mathbf{ c } }_{ AR_{ C } }$ is equal to $2^{ m }$ $\ket{ \mathbf{ a } }_{ AR }$ $\ket{ \mathbf{ b } }_{ AR_{ B } }$ $\ket{ \mathbf{ c } }_{ AR_{ C } }$, whereas if $\mathbf{ a } \oplus \mathbf{ b } \oplus \mathbf{ c } \oplus \mathbf{ s } \neq \mathbf{ 0 }$, the sum vanishes. Hence, expression \eqref{eq: $3$PSQDC Part 2 Expansion 1} can be simplified as

\begin{align}
	\label{eq: $3$PSQDC Part 2 Expansion 2}
	\ket{ \psi_{ 2 } }
	=
	\frac { 1 } { 2^{ m } }
	\sum_{ \mathbf{ a } \in \mathbb{ B }^{ m } }
	\sum_{ \mathbf{ b } \in \mathbb{ B }^{ m } }
	\sum_{ \mathbf{ c } \in \mathbb{ B }^{ m } }
	\
	\ket{ \mathbf{ a } }_{ AR }
	\
	\ket{ \mathbf{ b } }_{ AR_{ B } }
	\
	\ket{ \mathbf{ c } }_{ AR_{ C } }
	\ ,
	\quad \text{where} \quad
	\mathbf{ a } \oplus \mathbf{ b } \oplus \mathbf{ c } = \mathbf{ s }
	\ .
\end{align}

It is expedient to express the above equation in terms of individual bits. In view of \eqref{eq: Secret Bit Vector}, and the facts that $\mathbf{ a } = a_{ m - 1 } \dots a_{ 0 }$, $\mathbf{ b } = b_{ m - 1 } \dots b_{ 0 }$, and $\mathbf{ c } = c_{ m - 1 } \dots c_{ 0 }$, we can state the bitwise version of \eqref{eq: $3$ Player Fundamental Correlation Property}:

\begin{align}
	\label{eq: $3$ Player Bitwise Fundamental Correlation Property}
	\left
	\{
	\
	\begin{aligned}
		a_{ m - 1 }
		\oplus
		b_{ m - 1 }
		&\oplus
		c_{ m - 1 }
		=
		s_{ m - 1 }
		\\
		&\dots
		\\
		a_{ 0 }
		\oplus
		b_{ 0 }
		&\oplus
		c_{ 0 }
		=
		s_{ 0 }
	\end{aligned}
	\
	\right
	\}
	\ .
\end{align}

The above formulae convey the correlation of the contents of the registers $AR$, $AR_{ B }$, and $AR_{ C }$. This is a result of the entanglement between Alice, Bob and Charlie's registers in the initial state of the quantum circuit of Figure \ref{fig: $3$PSQDC Protocol Part 1}. Conceptually, we may consider this situation as follows: the contents of any two of the three registers may vary independently of each other, but then the contents of the remaining third register are completed defined by \eqref{eq: $3$ Player Fundamental Correlation Property} and its bitwise version \eqref{eq: $3$ Player Bitwise Fundamental Correlation Property}. These relations are referred to as the $3$ player \textbf{Hadamard Entanglement Property} and the $3$ player \textbf{Bitwise Hadamard Entanglement Property}, respectively.

Taking into account the effect of CNOT gates, as expressed by \eqref{eq: CNOT Gate}, we see that the action of the first group of $m$ CNOT gates, where each of the $m$ qubits in $AR$ serves as a control qubit targeting the corresponding qubit in $AR_{ B }$, results in

\begin{align}
	\label{eq: $3$PSQDC Part 2 Decryption 1}
	&
	\
	\left(
	{ \rm CNOT }
	\ket{ a_{ m - 1 } }
	\ket{ b_{ m - 1 } }
	\right)
	\
	\ket{ c_{ m - 1 } }
	\dots
	\left(
	{ \rm CNOT }
	\ket{ a_{ 0 } }
	\ket{ b_{ 0 } }
	\right)
	\
	\ket{ c_{ 0 } }
	\nonumber
	\\
	=
	&
	\
	\ket{ a_{ m - 1 } }
	\ket{ a_{ m - 1 } \oplus b_{ m - 1 } }
	\ket{ c_{ m - 1 } }
	\dots
	\ket{ a_{ 0 } }
	\ket{ a_{ 0 } \oplus b_{ 0 } }
	\ket{ c_{ 0 } }
	\nonumber
	\\
	=
	&
	\
	\ket{ \mathbf{ a } }_{ AR }
	\
	\ket{ \mathbf{ a } \oplus \mathbf{ b } }_{ AR_{ B } }
	\
	\ket{ \mathbf{ c } }_{ AR_{ C } }
	\ .
\end{align}

The subsequent action of the second group of $m$ CNOT gates, where each of the $m$ qubits in $AR_{ B }$ serves as a control qubit targeting the corresponding qubit in $AR_{ C }$, drives the circuit into the final state $\ket{ \psi_{ f } }$

\begin{align}
	\label{eq: $3$PSQDC Part 2 Decryption 2}
	&
	\
	\ket{ a_{ m - 1 } }
	\
	\left(
	{ \rm CNOT }
	\ket{ a_{ m - 1 } \oplus b_{ m - 1 } }
	\ket{ c_{ m - 1 } }
	\right)
	\dots
	\ket{ a_{ 0 } }
	\
	\left(
	{ \rm CNOT }
	\ket{ a_{ 0 } \oplus b_{ 0 } }
	\ket{ c_{ 0 } }
	\right)
	\nonumber
	\\
	=
	\hspace{ 0.100 cm }
	&
	\
	\ket{ a_{ m - 1 } }
	\ket{ b_{ m - 1 } \oplus c_{ m - 1 } }
	\ket{ a_{ m - 1 } \oplus b_{ m - 1 } \oplus c_{ m - 1 } }
	\dots
	\ket{ a_{ 0 } }
	\ket{ b_{ 0 } \oplus c_{ 0 } }
	\ket{ a_{ 0 } \oplus b_{ 0 } \oplus c_{ 0 } }
	\nonumber
	\\
	\overset { \eqref{eq: $3$ Player Bitwise Fundamental Correlation Property} } { = }
	&
	\
	\ket{ \mathbf{ a } }_{ AR }
	\
	\ket{ \mathbf{ a } \oplus \mathbf{ b } }_{ AR_{ B } }
	\
	\ket{ \mathbf{ s } }_{ AR_{ C } }
	\ .
\end{align}

Thus, in view of equations \eqref{eq: $3$PSQDC Part 2 Expansion 2} and \eqref{eq: $3$PSQDC Part 2 Decryption 2}) state $\ket{ \psi_{ f } }$ can be written as

\begin{align}
	\label{eq: $3$PSQDC Part 2 Decryption 3}
	\ket{ \psi_{ 3 } }
	=
	\frac { 1 } { 2^{ m } }
	\sum_{ \mathbf{ a } \in \mathbb{ B }^{ m } }
	\sum_{ \mathbf{ b } \in \mathbb{ B }^{ m } }
	\
	\ket{ \mathbf{ a } }_{ AR }
	\
	\ket{ \mathbf{ a } \oplus \mathbf{ b } }_{ AR_{ B } }
	\
	\ket{ \mathbf{ s } }_{ AR_{ C } }
	\ .
\end{align}

Now, Alice measures (in the computational basis) the contents of the register $AR_{ C }$ and obtains the secret bit vector $\mathbf{ s }$.

\section{Security analysis} \label{sec: Security Analysis}

The current section presents a unified security analysis of the $2$PSQDC and $3$PSQDC protocols. The ubiquitous Eve, as usual, undertakes the role of the adversary aiming to sabotage the protocol and steal the secret. We also assume the availability of pairwise classical authenticated channel for the purposes of implementing the eavesdropping detection and entanglement validation tests.
The advantage that quantum protocols exhibit over classical ones is that communication through the quantum channel involves an array of unique features, such as the no-cloning theorem \cite{wootters1982single}, the monogamy of entanglement \cite{coffman2000distributed}, and nonlocality \cite{brunner2014bell}, that can be used to inhibit Eve. For a recent comprehensive text analyzing security issues of quantum protocols in general we refer to \cite{Wolf2021} and the more recent \cite{Renner2023}. Extensive security analysis specifically for QSDC can be found in the thorough and very recent \cite{Pan2023} and \cite{Pan2024}.


We analyze the $3$ players case, involving Alice, Bob and Charlie, since the $2$ player case can be viewed as a special case. The setting now includes, in addition to our three protagonists, a forth notorious entity, traditionally named Eve, whose sole purpose is to devise and implement attacks against our protocol, aiming to acquire a piece of the secret information, or even, the complete secret information. Ultimately the security analysis of any quantum protocol rests on certain well-understood assumptions. For the sake of completeness, we briefly mention them at this point. First, we assume that quantum theory is correct, which in turn means that hallmark features such as the no-cloning theorem \cite{wootters1982single}, the monogamy of entanglement \cite{coffman2000distributed}, and nonlocality \cite{brunner2014bell} are valid. Clearly, if quantum protocols did not exhibit these properties, they would be useless. Secondly, we assume that quantum theory is complete, which implies that Eve is constrained by the laws of physics, and she cannot derive more information beyond what is predicted by quantum mechanics.

\subsection{Secret embedding security} \label{subsec: Secret Embedding Security}

The secret embedding phase begins only after the eavesdrop detection and entanglement validation tests have been successfully completed.

%

Let us be clear that even if Eve has successfully eavesdropped during the previous entanglement distribution phase, she will get no information whatsoever because no information has been encoded yet. However it is still possible that she may still disrupt the execution of the protocol. The probability of achieving that without being detected is practically zero. Let us consider what Eve may do during the distribution phase.

\begin{enumerate}
	\renewcommand\labelenumi{(\textbf{A}$_\theenumi$)}
	\item	\textbf{Measure and Resend}. Eve intercepts two qubits from each $\ket{ GHZ_{ 3 } }$ triplet during their transmission from Alice to Bob and Charlie, measures them and resends them back to Bob and Charlie. Eve will fail to discover any information because at this phase the $\ket{ GHZ_{ 3 } }$ triplets do not be carry any information. By the acts of measurement, Eve destroys the entanglement, and
	will be reveled during the entanglement validation test.
	\item	\textbf{Intercept and Resend fake $\ket{ GHZ_{ 3 } }$ triplets}. Eve intercepts two qubits from each $\ket{ GHZ_{ 3 } }$ triplet during their transmission from Alice to Bob and Charlie. Although cloning is prohibited by the no-cloning theorem, if Eve has already created a sufficient number of her own $\ket{ GHZ_{ 3 } }$ triplets, she may store the intercepted qubits, and, in their place, forward her own qubits. By employing such a strategy she will gain no information because at this phase the $\ket{ GHZ_{3} }$ triplets carry no information. Moreover, the fact that Eve knows nothing regarding the decoys, will again lead to
	erroneous outcomes when Bod and Charlie measure the decoys during the eavesdropping detection test.
\end{enumerate}

\subsection{Secret decryption security} \label{subsec: Secret Decryption Security}

Alice initiates the secret decryption phase only after the eavesdrop detection and entanglement validation tests have been successfully completed.

Let us now consider what attacks might Eve devise during the second quantum transmission, where Bob and Charlie each send $m$ qubits to Alice.

\begin{enumerate}
	\renewcommand\labelenumi{(\textbf{A}$_\theenumi$)}
	\item	\textbf{Measure and Resend}. Eve intercepts two qubits from each $\ket{ GHZ_{ 3 } }$ triplet during their transmission from Bob and Charlie to Alice, measures them and resends them back to Alice. Eve will fail to discover any information because she has no access to Alice's registers. By the acts of measurement, Eve destroys the entanglement. This random collapse of entanglement will result to
	erroneous outcomes that will be reveled during the entanglement verification check.
	\item	\textbf{Intercept and Resend fake $\ket{ GHZ_{ 3 } }$ triplets}. Eve intercepts two qubits from each $\ket{ GHZ_{ 3 } }$ triplet during their transmission from Bob and Charlie to Alice. Although cloning is prohibited by the no-cloning theorem, if Eve has already created a sufficient number of her own $\ket{ GHZ_{ 3 } }$ triplets, she may store the intercepted qubits, and, in their place, forward her own qubits. The flaw in this scenario is that Eve's entangled qubits do not carry the information embedded in Bob and Charlie's qubits. By employing such a strategy she will gain no information because she lacks the third piece of information existing in Alice's register. Obviously, the newly transmitted qubits will not contain the information Alice requires to reveal the correct $\mathbf{ s }$.
	Moreover, the fact that Eve knows nothing regarding the decoys, will again lead to
	wrong outcomes during the eavesdropping detection test.
	\item	\textbf{Entangle and Measure}. Once again, Eve intercepts two qubits from each $\ket{ GHZ_{ 3 } }$ triplet during their transmission from Bob and Charlie to Alice. This time Eve does not measure them, but entangles them with her ancilla state, and then sends the corresponding GHZ qubits to Alice. Eve waits until the protocol completes before measuring her qubits, hoping to gain useful information. However, the result of Eve's actions is that, instead of having $m$ $\ket{ GHZ_{ 3 } }$ triplets evenly distributed among Alice, Bob and Charlie, to end up with $m$ $\ket{ GHZ_{ 4 } }$ quadruples evenly distributed among Alice, Bob, Charlie, and Eve. Ultimately, Alice will derive an incorrect $\mathbf{ s }$, a fact that she will realize the eavesdropping detection check. Eve will also fail gain information about the correct $\mathbf{ s }$ because that would require the contents of Alice's register. Therefore, in this case too, Eve will fail, whereas Alice will be able to infer that Eve tempered with the protocol.
	\item	\textbf{PNS}. The photon number splitting attack (PNS), introduced in \cite{huttner1995quantum} and subsequently analyzed in \cite{lutkenhaus2000security, brassard2000limitations}, is regarded as one of the most effective attack strategies that Eve can employ against any quantum protocol. This attack exploits the fact that, due to technological limitations, photon sources occasionally do not emit single-photon signals, but may produce multiple identical photons instead of just one. This allows Eve to intercept pulses emanating from Alice for the distribution of the $\ket{ GHZ_{ 3 } }$ triplets, keep one photon from the multi-photon pulse for herself and send the remaining photons to Bob and Charlie without being detected during the transmission phase. As far as the $3$PSQDC protocol is concerned, this case resembles the Entangle and Measure attack analyzed above. Again, instead of $\ket{ GHZ_{ 3 } }$ triplets evenly distributed among Alice, Bob and Charlie, in reality there are $\ket{ GHZ_{ 4 } }$ quadruples evenly distributed among Alice, Bob, Charlie, and Eve. Eve becomes effectively the fourth player but still is unable to gain any information about the other players' measurements for the same reasons as in the previous case.
\end{enumerate}

The above security analysis demonstrates that both the $2$PSQDC and $3$PSQDC protocols are information-theoretically secure.

\section{Discussion and conclusions} \label{sec: Discussion and Conclusions}

In this work, we have introduced two new protocols for quantum secure direct communication. The first, called $2$PSQDC, involves information exchange between two entities, Alice and Bob. Subsequently, the $2$PSQDC was generalized in a intuitive and straightforward manner, so as to provide for quantum secure direct communication among three entities. The resulting protocol, which is called $3$PSQDC, can be seamlessly generalized to an arbitrary number of entities. Both protocols, which are proven to be information-theoretically secure, use the same idea, i.e., embedding the secret information into the entangled global state of the compound system via a unitary transform that uses the inner product modulo $2$ operation. This way of encoding the information is the main novelty of this paper that distinguishes it from the many previous works in the field. The advantage of this method is that it is seamlessly extensible and can be generalized to a setting involving three, or even more, players, as demonstrated with the $3$PSQDC protocol. This last case, is not only useful, but often necessary, when two spatially separated players posses only part the secret information that must be combined and transmitted to Alice in order for her to reveal the complete secret. Using the $3$PSQDC protocol, this task can be achieved in one go, without the need to apply a typical QSDC protocol twice, where Alice first receives Bob's information and afterwards Charlie's information. Assuming sufficient resources, the $3$PSQDC protocol can be extended in the obvious manner to allow for $n - 1$ players to simultaneously send information to Alice. It is also worth mentioning that both protocols are practically accessible, since they relay on EPR pairs, in the two player case, and $\ket{ GHZ_{ 3 } }$ triples, in the three player case, that can be generated with our current technology. Moreover, the local quantum circuits are characterized by uniformity and symmetry, being similar or identical, and are easily constructible as they employ only Hadamard and CNOT gates.

\bibliographystyle{ieeetr}
\bibliography{2-3PSQDC}

\begin{thebibliography}{10}

\bibitem{Shor1994}
P.~Shor, ``Algorithms for quantum computation: discrete logarithms and
  factoring,'' in {\em Proceedings 35th Annual Symposium on Foundations of
  Computer Science}, {IEEE} Comput. Soc. Press, 1994.

\bibitem{Grover1996}
L.~Grover, ``A fast quantum mechanical algorithm for database search,'' in {\em
  Proc. of the Twenty-Eighth Annual ACM Symposium on the Theory of Computing},
  1996.

\bibitem{IBMEagle2021}
J.~Chow, O.~Dial, and J.~Gambetta, ``{IBM} {Quantum} breaks the 100-qubit
  processor barrier.''
  \url{https://www.ibm.com/quantum/blog/127-qubit-quantum-processor-eagle},
  2021.
\newblock Accessed: 2024-03-02.

\bibitem{IBMOsprey2022}
I.~Newsroom, ``{IBM} unveils 400 qubit-plus quantum processor.''
  \url{https://newsroom.ibm.com/2022-11-09-IBM-Unveils-400-Qubit-Plus-Quantum-Processor-and-Next-Generation-IBM-Quantum-System-Two},
  2022.
\newblock Accessed: 2024-03-02.

\bibitem{IBMCondor2023}
J.~Gambetta, ``The hardware and software for the era of quantum utility is
  here.'' \url{https://www.ibm.com/quantum/blog/quantum-roadmap-2033}, 2023.
\newblock Accessed: 2024-03-02.

\bibitem{IBMHeron2024}
I.~Newsroom, ``{IBM} launches its most advanced quantum computers, fueling new
  scientific value and progress towards quantum advantage.''
  \url{https://newsroom.ibm.com/2024-11-13-ibm-launches-its-most-advanced-quantum-computers,-fueling-new-scientific-value-and-progress-towards-quantum-advantage},
  2024.
\newblock Accessed: 2024-11-21.

\bibitem{chen2016report}
L.~Chen, L.~Chen, S.~Jordan, Y.-K. Liu, D.~Moody, R.~Peralta, R.~Perlner, and
  D.~Smith-Tone, {\em Report on post-quantum cryptography}, vol.~12.
\newblock US Department of Commerce, National Institute of Standards and
  Technology, 2016.

\bibitem{alagic2019status}
G.~Alagic, G.~Alagic, J.~Alperin-Sheriff, D.~Apon, D.~Cooper, Q.~Dang, Y.-K.
  Liu, C.~Miller, D.~Moody, R.~Peralta, {\em et~al.}, {\em Status report on the
  first round of the NIST post-quantum cryptography standardization process}.
\newblock US Department of Commerce, National Institute of Standards and
  Technology~…, 2019.

\bibitem{alagic2020status}
G.~Alagic, J.~Alperin-Sheriff, D.~Apon, D.~Cooper, Q.~Dang, J.~Kelsey, Y.-K.
  Liu, C.~Miller, D.~Moody, R.~Peralta, {\em et~al.}, ``Status report on the
  second round of the nist post-quantum cryptography standardization process,''
  {\em US Department of Commerce, NIST}, 2020.

\bibitem{alagic2022status}
G.~Alagic, D.~Apon, D.~Cooper, Q.~Dang, T.~Dang, J.~Kelsey, J.~Lichtinger,
  C.~Miller, D.~Moody, R.~Peralta, {\em et~al.}, ``Status report on the third
  round of the nist post-quantum cryptography standardization process,'' {\em
  National Institute of Standards and Technology, Gaithersburg}, 2022.

\bibitem{Bennett1984}
C.~H. Bennett and G.~Brassard, ``Quantum cryptography: Public key distribution
  and coin tossing,'' in {\em Proceedings of the IEEE International Conference
  on Computers, Systems, and Signal Processing}, pp.~175--179, {IEEE} Computer
  Society Press, 1984.

\bibitem{Ekert1991}
A.~K. Ekert, ``Quantum cryptography based on bell's theorem,'' {\em Physical
  Review Letters}, vol.~67, no.~6, pp.~661--663, 1991.

\bibitem{Gisin2004}
N.~Gisin, G.~Ribordy, H.~Zbinden, D.~Stucki, N.~Brunner, and V.~Scarani,
  ``Towards practical and fast quantum cryptography,'' {\em arXiv preprint
  quant-ph/0411022}, 2004.

\bibitem{inoue2002differential}
K.~Inoue, E.~Waks, and Y.~Yamamoto, ``Differential phase shift quantum key
  distribution,'' {\em Physical review letters}, vol.~89, no.~3, p.~037902,
  2002.

\bibitem{guan2015experimental}
J.-Y. Guan, Z.~Cao, Y.~Liu, G.-L. Shen-Tu, J.~S. Pelc, M.~Fejer, C.-Z. Peng,
  X.~Ma, Q.~Zhang, and J.-W. Pan, ``Experimental passive round-robin
  differential phase-shift quantum key distribution,'' {\em Physical review
  letters}, vol.~114, no.~18, p.~180502, 2015.

\bibitem{waks2006security}
E.~Waks, H.~Takesue, and Y.~Yamamoto, ``Security of differential-phase-shift
  quantum key distribution against individual attacks,'' {\em Physical Review
  A}, vol.~73, no.~1, p.~012344, 2006.

\bibitem{Ampatzis2021}
M.~Ampatzis and T.~Andronikos, ``{QKD} based on symmetric entangled
  bernstein-vazirani,'' {\em Entropy}, vol.~23, no.~7, p.~870, 2021.

\bibitem{Hillery1999}
M.~Hillery, V.~Bu{\v{z}}ek, and A.~Berthiaume, ``Quantum secret sharing,'' {\em
  Physical Review A}, vol.~59, no.~3, p.~1829, 1999.

\bibitem{Ampatzis2022}
M.~Ampatzis and T.~Andronikos, ``A symmetric extensible protocol for quantum
  secret sharing,'' {\em Symmetry}, vol.~14, no.~8, p.~1692, 2022.

\bibitem{Ampatzis2023}
M.~Ampatzis and T.~Andronikos, ``Quantum secret aggregation utilizing a network
  of agents,'' {\em Cryptography}, vol.~7, no.~1, p.~5, 2023.

\bibitem{Andronikos2024b}
T.~Andronikos, ``A distributed and parallel (k, n) qss scheme with verification
  capability,'' {\em Mathematics}, vol.~12, no.~23, p.~3782, 2024.

\bibitem{Bennett1993}
C.~H. Bennett, G.~Brassard, C.~Cr{\'{e}}peau, R.~Jozsa, A.~Peres, and W.~K.
  Wootters, ``Teleporting an unknown quantum state via dual classical and
  einstein-podolsky-rosen channels,'' {\em Physical Review Letters}, vol.~70,
  no.~13, pp.~1895--1899, 1993.

\bibitem{attasena2017secret}
V.~Attasena, J.~Darmont, and N.~Harbi, ``Secret sharing for cloud data
  security: a survey,'' {\em The VLDB Journal}, vol.~26, no.~5, pp.~657--681,
  2017.

\bibitem{ermakova2013secret}
T.~Ermakova and B.~Fabian, ``Secret sharing for health data in multi-provider
  clouds,'' in {\em 2013 IEEE 15th conference on business informatics},
  pp.~93--100, IEEE, 2013.

\bibitem{cha2021blockchain}
J.~Cha, S.~K. Singh, T.~W. Kim, and J.~H. Park, ``Blockchain-empowered cloud
  architecture based on secret sharing for smart city,'' {\em Journal of
  Information Security and Applications}, vol.~57, p.~102686, 2021.

\bibitem{Sun2020}
X.~Sun, P.~Kulicki, and M.~Sopek, ``Multi-party quantum byzantine agreement
  without entanglement,'' {\em Entropy}, vol.~22, no.~10, p.~1152, 2020.

\bibitem{Qu2023}
Z.~Qu, Z.~Zhang, B.~Liu, P.~Tiwari, X.~Ning, and K.~Muhammad, ``Quantum
  detectable byzantine agreement for distributed data trust management in
  blockchain,'' {\em Information Sciences}, vol.~637, p.~118909, 2023.

\bibitem{Photonic2024}
Photonic, ``Photonic demonstrates distributed entanglement between modules,
  marking significant milestone toward scalable quantum computing and
  networking.''
  \url{https://photonic.com/news/photonic-demonstrates-distributed-entanglement-between-modules/},
  2024.
\newblock Accessed: 2024-11-21.

\bibitem{NuQuantum2024}
N.~Quantum, ``Announcing the qubit-photon interface (qpi): towards unlocking
  modular and scalable distributed quantum computing.''
  \url{https://www.nu-quantum.com/news/qubit-photon-interface-qpi-towards-unlocking-modular-and-scalable-distributed-quantum-computing},
  2024.
\newblock Accessed: 2024-11-21.

\bibitem{Cacciapuoti2024}
A.~S. Cacciapuoti, J.~Illiano, M.~Viscardi, and M.~Caleffi, ``Multipartite
  entanglement distribution in the quantum internet: Knowing when to stop!,''
  {\em IEEE Transactions on Network and Service Management}, pp.~1--1, 2024.

\bibitem{Illiano2024}
J.~Illiano, M.~Caleffi, M.~Viscardi, and A.~S. Cacciapuoti, ``Quantum mac:
  Genuine entanglement access control via many-body dicke states,'' {\em IEEE
  Transactions on Communications}, vol.~72, no.~4, pp.~2090--2105, 2024.

\bibitem{Long2002}
G.~L. Long and X.~S. Liu, ``Theoretically efficient high-capacity
  quantum-key-distribution scheme,'' {\em Physical Review A}, vol.~65, no.~3,
  p.~032302, 2002.

\bibitem{Deng2003}
F.-G. Deng, G.~L. Long, and X.-S. Liu, ``Two-step quantum direct communication
  protocol using the einstein-podolsky-rosen pair block,'' {\em Physical Review
  A}, vol.~68, no.~4, p.~042317, 2003.

\bibitem{Deng2004}
F.-G. Deng and G.~L. Long, ``Secure direct communication with a quantum
  one-time pad,'' {\em Physical Review A}, vol.~69, no.~5, p.~052319, 2004.

\bibitem{Wang2005}
C.~Wang, F.-G. Deng, Y.-S. Li, X.-S. Liu, and G.~L. Long, ``Quantum secure
  direct communication with high-dimension quantum superdense coding,'' {\em
  Physical Review A}, vol.~71, no.~4, p.~044305, 2005.

\bibitem{Wang2005a}
C.~Wang, F.~G. Deng, and G.~L. Long, ``Multi-step quantum secure direct
  communication using multi-particle
  green{\textendash}horne{\textendash}zeilinger state,'' {\em Optics
  Communications}, vol.~253, no.~1-3, pp.~15--20, 2005.

\bibitem{Pan2023}
D.~Pan, X.-T. Song, and G.-L. Long, ``Free-space quantum secure direct
  communication: Basics, progress, and outlook,'' {\em Advanced Devices \&
  Instrumentation}, vol.~4, 2023.

\bibitem{Pan2024}
D.~Pan, G.-L. Long, L.~Yin, Y.-B. Sheng, D.~Ruan, S.~X. Ng, J.~Lu, and
  L.~Hanzo, ``The evolution of quantum secure direct communication: On the road
  to the qinternet,'' {\em IEEE Communications Surveys \& Tutorials}, vol.~26,
  no.~3, pp.~1898--1949, 2024.

\bibitem{Meyer1999}
D.~A. Meyer, ``Quantum strategies,'' {\em Physical Review Letters}, vol.~82,
  no.~5, p.~1052, 1999.

\bibitem{Eisert1999}
J.~Eisert, M.~Wilkens, and M.~Lewenstein, ``Quantum games and quantum
  strategies,'' {\em Physical Review Letters}, vol.~83, no.~15, p.~3077, 1999.

\bibitem{Andronikos2018}
T.~Andronikos, A.~Sirokofskich, K.~Kastampolidou, M.~Varvouzou, K.~Giannakis,
  and A.~Singh, ``Finite automata capturing winning sequences for all possible
  variants of the {PQ} penny flip game,'' {\em Mathematics}, vol.~6, p.~20, Feb
  2018.

\bibitem{Andronikos2021}
T.~Andronikos and A.~Sirokofskich, ``The connection between the {PQ} penny flip
  game and the dihedral groups,'' {\em Mathematics}, vol.~9, no.~10, p.~1115,
  2021.

\bibitem{Andronikos2022a}
T.~Andronikos, ``Conditions that enable a player to surely win in sequential
  quantum games,'' {\em Quantum Information Processing}, vol.~21, no.~7, 2022.

\bibitem{Giannakis2015a}
K.~Giannakis, C.~Papalitsas, K.~Kastampolidou, A.~Singh, and T.~Andronikos,
  ``Dominant strategies of quantum games on quantum periodic automata,'' {\em
  Computation}, vol.~3, pp.~586--599, nov 2015.

\bibitem{Koh2024}
D.~E. Koh, K.~Kumar, and S.~T. Goh, ``Quantum volunteer's dilemma,'' 2024.

\bibitem{Andronikos2022}
T.~Andronikos and M.~Stefanidakis, ``A two-party quantum parliament,'' {\em
  Algorithms}, vol.~15, no.~2, p.~62, 2022.

\bibitem{Theocharopoulou2019}
G.~Theocharopoulou, K.~Giannakis, C.~Papalitsas, S.~Fanarioti, and
  T.~Andronikos, ``Elements of game theory in a bio-inspired model of
  computation,'' in {\em 2019 10th International Conference on Information,
  Intelligence, Systems and Applications ({IISA})}, pp.~1--4, {IEEE}, jul 2019.

\bibitem{Kastampolidou2020a}
K.~Kastampolidou, M.~N. Nikiforos, and T.~Andronikos, ``A brief survey of the
  prisoners' dilemma game and its potential use in biology,'' in {\em Advances
  in Experimental Medicine and Biology}, pp.~315--322, Springer International
  Publishing, 2020.

\bibitem{Kostadimas2021}
D.~Kostadimas, K.~Kastampolidou, and T.~Andronikos, ``Correlation of biological
  and computer viruses through evolutionary game theory,'' in {\em 2021 16th
  International Workshop on Semantic and Social Media Adaptation {\&}
  Personalization ({SMAP})}, {IEEE}, 2021.

\bibitem{Kastampolidou2020}
K.~Kastampolidou and T.~Andronikos, ``A survey of evolutionary games in
  biology,'' in {\em Advances in Experimental Medicine and Biology},
  pp.~253--261, Springer International Publishing, 2020.

\bibitem{Kastampolidou2021}
K.~Kastampolidou and T.~Andronikos, ``Microbes and the games they play,'' in
  {\em {GeNeDis} 2020}, pp.~265--271, Springer International Publishing, 2021.

\bibitem{Papalitsas2021}
C.~Papalitsas, K.~Kastampolidou, and T.~Andronikos, ``Nature and
  quantum-inspired procedures {\textendash} a short literature review,'' in
  {\em {GeNeDis} 2020}, pp.~129--133, Springer International Publishing, 2021.

\bibitem{Kastampolidou2023}
K.~Kastampolidou and T.~Andronikos, ``Game theory and other unconventional
  approaches to biological systems,'' in {\em Handbook of Computational
  Neurodegeneration}, pp.~163--180, Springer International Publishing, 2023.

\bibitem{Adam2023}
S.~Adam, P.~Karastathis, D.~Kostadimas, K.~Kastampolidou, and T.~Andronikos,
  ``Protein misfolding and neurodegenerative diseases: A game theory
  perspective,'' in {\em Handbook of Computational Neurodegeneration},
  pp.~863--874, Springer International Publishing, 2023.

\bibitem{Nielsen2010}
M.~A. Nielsen and I.~L. Chuang, {\em Quantum computation and quantum
  information}.
\newblock Cambridge University Press, 2010.

\bibitem{Cruz2019}
D.~Cruz, R.~Fournier, F.~Gremion, A.~Jeannerot, K.~Komagata, T.~Tosic,
  J.~Thiesbrummel, C.~L. Chan, N.~Macris, M.-A. Dupertuis, and
  C.~Javerzac-Galy, ``Efficient quantum algorithms for {GHZ} and w states, and
  implementation on the {IBM} quantum computer,'' {\em Advanced Quantum
  Technologies}, vol.~2, no.~5-6, p.~1900015, 2019.

\bibitem{Mermin2007}
N.~Mermin, {\em Quantum Computer Science: An Introduction}.
\newblock Cambridge University Press, 2007.

\bibitem{Wong2022}
T.~G. Wong, {\em Introduction to classical and quantum computing}.
\newblock Rooted Grove, 2022.

\bibitem{Andronikos2023b}
T.~Andronikos and A.~Sirokofskich, ``One-to-many simultaneous secure quantum
  information transmission,'' {\em Cryptography}, vol.~7, no.~4, p.~64, 2023.

\bibitem{wootters1982single}
W.~K. Wootters and W.~H. Zurek, ``A single quantum cannot be cloned,'' {\em
  Nature}, vol.~299, no.~5886, pp.~802--803, 1982.

\bibitem{coffman2000distributed}
V.~Coffman, J.~Kundu, and W.~K. Wootters, ``Distributed entanglement,'' {\em
  Physical Review A}, vol.~61, no.~5, p.~052306, 2000.

\bibitem{brunner2014bell}
N.~Brunner, D.~Cavalcanti, S.~Pironio, V.~Scarani, and S.~Wehner, ``Bell
  nonlocality,'' {\em Reviews of Modern Physics}, vol.~86, no.~2, p.~419, 2014.

\bibitem{Deng2008}
F.-G. Deng, X.-H. Li, and H.-Y. Zhou, ``Efficient high-capacity quantum secret
  sharing with two-photon entanglement,'' {\em Physics Letters A}, vol.~372,
  no.~12, pp.~1957--1962, 2008.

\bibitem{Yang2009}
Y.-G. Yang and Q.-Y. Wen, ``An efficient two-party quantum private comparison
  protocol with decoy photons and two-photon entanglement,'' {\em Journal of
  Physics A: Mathematical and Theoretical}, vol.~42, no.~5, p.~055305, 2009.

\bibitem{Tseng2011}
H.-Y. Tseng, J.~Lin, and T.~Hwang, ``New quantum private comparison protocol
  using epr pairs,'' {\em Quantum Information Processing}, vol.~11, no.~2,
  pp.~373--384, 2011.

\bibitem{Chang2013}
Y.-J. Chang, C.-W. Tsai, and T.~Hwang, ``Multi-user private comparison protocol
  using ghz class states,'' {\em Quantum Information Processing}, vol.~12,
  no.~2, pp.~1077--1088, 2013.

\bibitem{Hung2016}
S.-M. Hung, S.-L. Hwang, T.~Hwang, and S.-H. Kao, ``Multiparty quantum private
  comparison with almost dishonest third parties for strangers,'' {\em Quantum
  Information Processing}, vol.~16, no.~2, 2016.

\bibitem{Ye2018}
C.-Q. Ye and T.-Y. Ye, ``Multi-party quantum private comparison of size
  relation with d-level single-particle states,'' {\em Quantum Information
  Processing}, vol.~17, no.~10, 2018.

\bibitem{Wu2021}
W.~Wu and Y.~Zhao, ``Quantum private comparison of size using d-level bell
  states with a semi-honest third party,'' {\em Quantum Information
  Processing}, vol.~20, no.~4, 2021.

\bibitem{Hou2024}
M.~Hou and Y.~Wu, ``Single-photon-based quantum secure protocol for the
  socialist millionaires’ problem,'' {\em Frontiers in Physics}, vol.~12,
  2024.

\bibitem{Neigovzen2008}
R.~Neigovzen, C.~Rod{\'{o}}, G.~Adesso, and A.~Sanpera, ``Multipartite
  continuous-variable solution for the byzantine agreement problem,'' {\em
  Physical Review A}, vol.~77, no.~6, p.~062307, 2008.

\bibitem{Feng2019}
Y.~Feng, R.~Shi, J.~Zhou, Q.~Liao, and Y.~Guo, ``Quantum byzantine agreement
  with tripartite entangled states,'' {\em International Journal of Theoretical
  Physics}, vol.~58, no.~5, pp.~1482--1498, 2019.

\bibitem{Wang2022a}
W.~Wang, Y.~Yu, and L.~Du, ``Quantum blockchain based on asymmetric quantum
  encryption and a stake vote consensus algorithm,'' {\em Scientific Reports},
  vol.~12, no.~1, 2022.

\bibitem{Yang2022}
Z.~Yang, T.~Salman, R.~Jain, and R.~D. Pietro, ``Decentralization using quantum
  blockchain: A theoretical analysis,'' {\em IEEE Transactions on Quantum
  Engineering}, vol.~3, pp.~1--16, 2022.

\bibitem{Ikeda2023c}
K.~Ikeda and A.~Lowe, ``Quantum protocol for decision making and verifying
  truthfulness among n‐quantum parties: Solution and extension of the quantum
  coin flipping game,'' {\em IET Quantum Communication}, vol.~4, no.~4,
  pp.~218--227, 2023.

\bibitem{Wolf2021}
R.~Wolf, {\em Quantum Key Distribution}.
\newblock Springer International Publishing, 2021.

\bibitem{Renner2023}
R.~Renner and R.~Wolf, ``Quantum advantage in cryptography,'' {\em {AIAA}
  Journal}, vol.~61, no.~5, pp.~1895--1910, 2023.

\bibitem{huttner1995quantum}
B.~Huttner, N.~Imoto, N.~Gisin, and T.~Mor, ``Quantum cryptography with
  coherent states,'' {\em Physical Review A}, vol.~51, no.~3, p.~1863, 1995.

\bibitem{lutkenhaus2000security}
N.~L{\"u}tkenhaus, ``Security against individual attacks for realistic quantum
  key distribution,'' {\em Physical Review A}, vol.~61, no.~5, p.~052304, 2000.

\bibitem{brassard2000limitations}
G.~Brassard, N.~L{\"u}tkenhaus, T.~Mor, and B.~C. Sanders, ``Limitations on
  practical quantum cryptography,'' {\em Physical review letters}, vol.~85,
  no.~6, p.~1330, 2000.

\end{thebibliography}

\end{document}